\documentclass[letterpaper]{article} % DO NOT CHANGE THIS
\usepackage{aaai24}  % DO NOT CHANGE THIS
\usepackage{times}  % DO NOT CHANGE THIS
\usepackage{helvet}  % DO NOT CHANGE THIS
\usepackage{courier}  % DO NOT CHANGE THIS
\usepackage[hyphens]{url}  % DO NOT CHANGE THIS
\usepackage{graphicx} % DO NOT CHANGE THIS
\usepackage{natbib}  % DO NOT CHANGE THIS AND DO NOT ADD ANY OPTIONS TO IT
\usepackage{caption} % DO NOT CHANGE THIS AND DO NOT ADD ANY OPTIONS TO IT
\usepackage{algorithm}
\usepackage{algorithmic}

\usepackage{newfloat}
\usepackage{listings}
\DeclareCaptionStyle{ruled}{labelfont=normalfont,labelsep=colon,strut=off} % DO NOT CHANGE THIS
\floatstyle{ruled}
\newfloat{listing}{tb}{lst}{}
\floatname{listing}{Listing}
\usepackage{xspace}
\newcommand{\omm}{\textsc{Omm}\xspace} 
\usepackage{amsfonts,amstext,amsmath,dsfont}
\usepackage[capitalise,nameinlink]{cleveref}
\usepackage[dvipsnames]{xcolor}
\usepackage{subfigure}
\usepackage{tabularx}
\usepackage{booktabs}

\usepackage[final]{pdfpages}

\usepackage{soul}
\usepackage{xspace}
\usepackage[colorinlistoftodos,textsize=tiny]{todonotes}

\definecolor{navy}{rgb}{0.1, 0.1, 0.8}
\definecolor[named]{gray}{rgb}{0.4, 0.4, 0.4}
\definecolor[named]{olive}{rgb}{0.1, 0.5, 0.1}
\definecolor[named]{ruby}{rgb}{0.8, 0.1, 0.3}
\definecolor{darkpastelgreen}{rgb}{0.01, 0.75, 0.24}
\definecolor{celestialblue}{rgb}{0.29, 0.59, 0.82}
\definecolor{coral}{rgb}{1.0, 0.5, 0.31}
\definecolor{Goldenrod}{rgb}{0.8,0.8,0}

\newcommand{\eat}[1]{}

\title{Opinion Market Model:\\ Stemming Far-Right Opinion Spread Using Positive Interventions}
\author{
    Pio Calderon, Rohit Ram, Marian-Andrei Rizoiu
}
\affiliations{
    University of Technology Sydney\\
    piogabrielle.b.calderon@student.uts.edu.au, rohit.ram@student.uts.edu.au, marian-andrei.rizoiu@uts.edu.au
}

\begin{document}

\maketitle

\begin{abstract}
    Online extremism has severe societal consequences, including normalizing hate speech, user radicalization, and increased social divisions. 
    Various mitigation strategies have been explored to address these consequences.
    One such strategy uses positive interventions: controlled signals that add attention to the opinion ecosystem to boost certain opinions. 
    To evaluate the effectiveness of positive interventions, we introduce the Opinion Market Model (\omm), a two-tier online opinion ecosystem model that considers both inter-opinion interactions and the role of positive interventions. 
    The size of the opinion attention market is modeled in the first tier using the multivariate discrete-time Hawkes process; 
    in the second tier, opinions cooperate and compete for market share, given limited attention using the market share attraction model. 
    We demonstrate the convergence of our proposed estimation scheme on a synthetic dataset. 
    Next, we test \omm on two learning tasks, applying to two real-world datasets to predict attention market shares and uncover latent relationships between online items.
    The first dataset comprises Facebook and Twitter discussions containing moderate and far-right opinions about bushfires and climate change. 
    The second dataset captures popular VEVO artists' YouTube and Twitter attention volumes.
    \omm outperforms the state-of-the-art predictive models on both datasets and captures latent cooperation-competition relations. 
    We uncover (1) self- and cross-reinforcement between far-right and moderate opinions on the bushfires and (2) pairwise artist relations that correlate with real-world interactions such as collaborations and long-lasting feuds.
    Lastly, we use \omm as a testbed for positive interventions and show how media coverage modulates the spread of far-right opinions.
\end{abstract}

\section{Introduction}

\begin{figure}[!ht]
    \centering
    \includegraphics[width=0.44\textwidth]{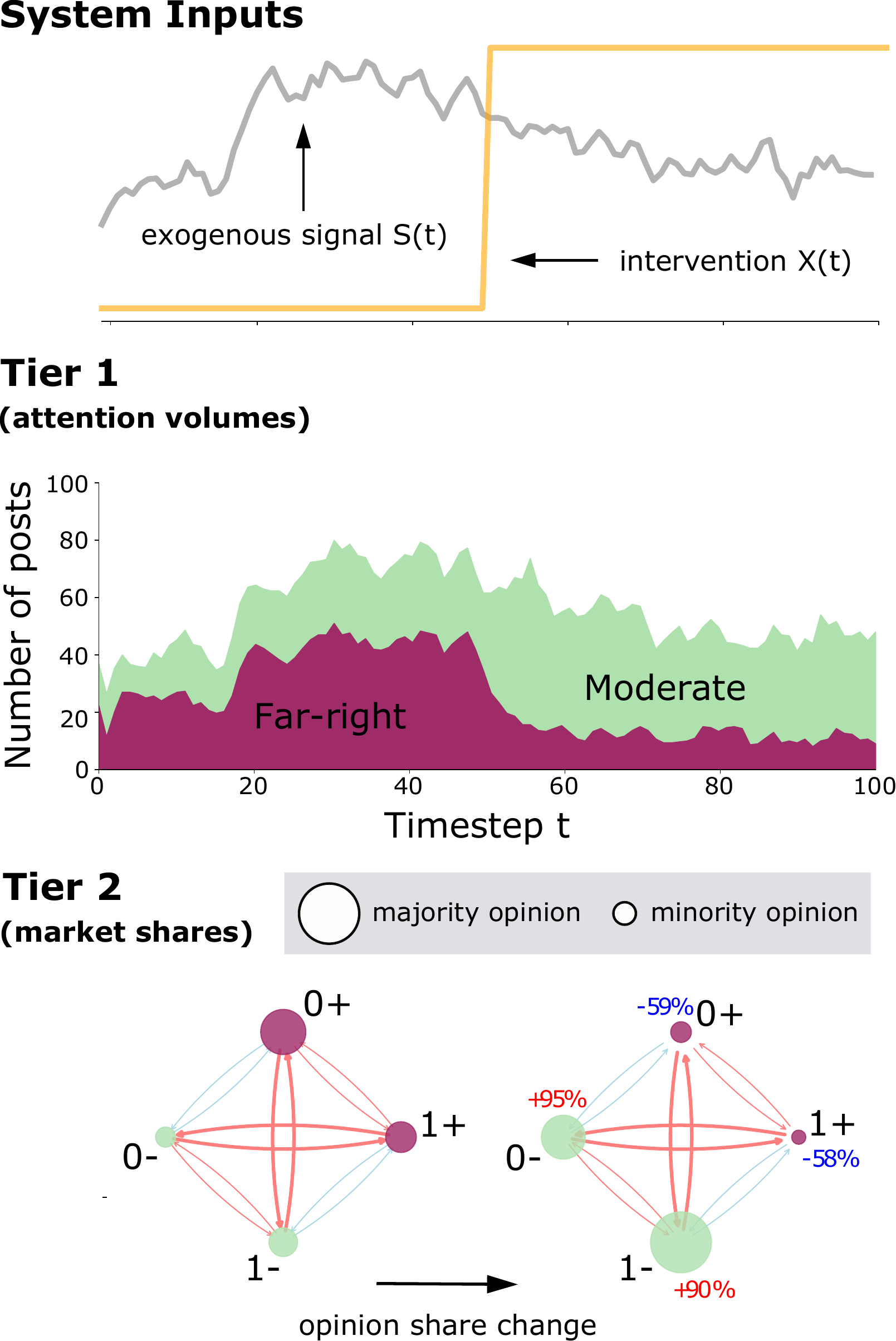}
    \caption{
        We illustrate how the positive intervention $X(t)$ (defined in \cref{eq:opinionsharemodel_tendency}) suppresses far-right opinions on a simulated toy opinion ecosystem with two far-right (0+, 1+) and two moderate (0-, 1-) opinions. 
        For instance, 0+ and 1+ can represent the opinions ``the Greens policies caused the Australian bushfires" and ``mainstream media cannot be trusted," respectively; 0- and 1- can be obtained as their negations.
        \textit{Top row}: the exogenous signal $S(t)$ (defined in \cref{eq:opinionvolumemodel}) and the intervention $X(t)$.
        %The inset shows the instantaneous reaction of the four opinions to the intervention (\textcolor{blue}{blue} signifies inhibition and \textcolor{red}{red} is reinforcement).
        \textit{Middle row}: total daily opinion market size quantified by our model's first tier, split into far-right (+) and moderate (-) opinion volumes.
        \textit{Bottom row}: market shares and the interactions between the four opinions estimated by our model's second tier. 
        Nodes are opinions; their sizes indicate market share; edges represent exciting (\textcolor{red}{red}) and inhibiting (\textcolor{blue}{blue}) relations.
        $X(t)$ suppresses far-right opinions for $t > 50$.
        Shown are average market shares before (left) and after (right) $t=50$.}
    % \caption{
    %     Suppressing far-right opinion with positive interventions.
    %     A simulated toy opinion ecosystem with two far-right (+) and two moderate (-) opinions.
    %     \textit{Bottom row}: the exogenous signal $S(t)$ (defined in \cref{eq:opinionvolumemodel}) and intervention $X(t)$ (defined in \cref{eq:opinionsharemodel_tendency}). 
    %     We set $X(t)$ as a step function with change point $t=50$. 
    %     The inset shows the instantaneous reaction of the four opinions to the intervention (\textcolor{blue}{blue} signifies inhibition and \textcolor{red}{red} is reinforcement).
    %     %  elasticities (see \cref{elasticx}) for the two far-right (+) and two moderate (-) opinions on the top row. 
    %     \textit{Middle row}: Total daily opinion market size quantified by our model's first tier, split into far-right (+) and moderate (-) opinion volumes.
    %     \textit{Top row}: Market shares and the interactions between the four opinions estimated by our model's second tier. 
    %     Nodes are opinions; their sizes indicate market share; edges represent exciting (\textcolor{red}{red}) and inhibiting (\textcolor{blue}{blue}) relations.
    %     $X(t)$ suppresses far-right opinions for $t > 50$.
    %     Shown are average market shares before (left) and after (right) $t=50$.}
	\label{fig:teaser}
\end{figure}

Online social media platforms are fertile grounds for deliberation and opinion formation \cite{Gupta2022,Upadhyay2019}.
Opinions thrive in the \textit{online opinion ecosystem}, interconnected online social platforms where they interact -- cooperating or competing for the finite public attention~\cite{Wu2019}. % -- within and across online social groups. 

We delineate two types of interventions to mitigate the spread of extremist views.
\textit{Negative interventions} aim to subtract attention from the opinion ecosystem by placing fact-check warnings on postings \cite{Nekmat2020}, shadowbanning \cite{Young2022} or outright banning extremist social media groups and accounts \cite{Jackson2019}.
While negative interventions are effective \cite{Clayton2020}, they are available solely to the social media platforms that tend to use them sparingly \cite{Porter2021}.

\textit{Positive interventions} \cite{PI2021}, such as misinformation debunking \cite{Haciyakupoglu2018,Shu2019} and increasing media coverage \cite{Horowitz2022}, mitigate extremist views by adding attention to the online opinion ecosystem through informing the public, redistributing attention away from extremist, and toward moderate views. 
Such interventions are typically in the hands of government and media agencies \cite{Radsch2016}. 
% These act in a different way to social media platform-enacted 
Testing the viability of positive interventions requires capturing reactions to interventions and inter-opinion interactions.

This work develops a model for the dynamics of the opinion ecosystem and a testbed for evaluating positive interventions.
We focus on two open questions. 
The first question explores the analogy between opinions and economic goods. 
In a competitive economic market of limited resources, coexisting goods can interact in one of two ways: either they compete for market share (\textit{substitute} brands, like Pepsi and Coke) or reinforce each other (\textit{complementary} items, like bread and butter). 
We argue that opinions in the online ecosystem behave similarly, allowing us to leverage market share modeling tools \cite{Cooper1993}. 
The first research question is: \textbf{Can we model the online opinion ecosystem as an environment where opinions cooperate or compete for market share?} 
We propose the Opinion Market Model\footnote{\label{note1}The code and datasets are available at https://github.com/behavioral-ds/opinion-market-model.} (\omm), a two-tier model to address this question.
\cref{fig:teaser} showcases a simple opinion ecosystem under intervention, with two opinions (denoted $0$ and $1$) on a single social media platform.
Each opinion has two polarities: far-right supporters (+) and moderate debunkers (-).
Exogenous signals (shown in gray in the top panel of \cref{fig:teaser}) and interventions (shown in yellow) modulate the dynamics of the opinions' sizes.
Exogenous signals are naturally occurring events like bushfires, floods, or political speeches.
Interventions (like increased media coverage) are designed to add attention to the opinion ecosystem, increasing the market share of certain opinions while suppressing others.
The first tier of \omm (middle row in \cref{fig:teaser}) uses a discrete-time Hawkes process to estimate the size of the opinion attention market -- that is, the daily number of postings featuring opinions. 
The Hawkes process has been widely used to model online attention \citep{Rizoiu2017,Zarezade2017} due to its ability to account for exogenous factors and the endogenous ``word-of-mouth'' through its self- and cross-exciting property.
The second tier of \omm (bottom row in \cref{fig:teaser}) leverages a market share attraction model to capture opinion interactions -- we assume that online opinions compete for the users' limited online attention \citep{Weng2012,Gelper2021}. 
For the example in \cref{fig:teaser}, opinions $0-$ and $1+$ have a strong reinforcing relation (shown as red arrows), while $1-$ and $1+$ have a weak competing relation (blue arrows).

We test \omm on two real-world datasets\textsuperscript{\ref{note1}}. 
The first contains Facebook and Twitter discussions expressing moderate and far-right opinions on bushfires and climate change \cite{Kong2021}. 
The second captures the YouTube views and Twitter mentions for the most popular VEVO artists' songs in 2017 \cite{Wu2019}.
We evaluate \omm on two tasks: predicting attention market share and exposing relationships between online items.
For the predictive task, \omm outperforms the current state of the art in market share modeling (Correlated Cascades \cite{Zarezade2017} and Competing Products \cite{Valera2015}) and predictive baselines on both datasets.
%  and demonstrates strong interpretation capabilities on , highlighting the generalisability of \omm in different application contexts.
For the second task,
we leverage the \omm to expose the relations between opinions on the two platforms.
For the bushfire case study, no significant interactions occurred on Facebook, as postings were collected from far-right public groups with limited interaction with the opposing side.
On Twitter, we observe self-reinforcement behavior of both far-right and moderate opinions, probably due to the \textit{echo chamber effect} \cite{Cinelli2021} -- reinforcing one's beliefs due to repeated interactions with users sharing similar ideologies on social platforms.
Surprisingly, we notice that opposing views reinforce each other, probably due to the deliberative nature of Twitter, where far-right sympathizers and opponents oppose each other. 
For the VEVO artists case study, we uncover nontrivial pairwise interactions of music artists correlating with real-world relationships -- such as Ariana Grande's and Calvin Harris' reinforcement relationship due to their collaboration ``Heatstroke'' and Taylor Swift's and Justin Bieber's inhibiting relationship.

Our second research question is: \textbf{Can we test the sensitivity of the opinion ecosystem to positive interventions?} 
\omm accounts for positive interventions -- controlled external signals to boost certain opinions.
In \cref{fig:teaser} an intervention is performed for $t > 50$, which suppresses the far-right opinions (+), leading to the shrinking of their market share.
We use \omm for two tasks: 
first, to estimate whether interventions effectively shape the opinion ecosystem and, second, to construct what-if scenarios as synthetic testbeds for future interventions.
For the bushfire case study, we test whether news coverage from reputable and controversial media outlets suppresses or aids the spread of far-right opinions.
We fit \omm twice: with and without media coverage.
We find a better fit with the intervention, suggesting that media coverage actively shapes the opinion ecosystem.
We perform synthetic what-if experiments: we vary the level of media coverage, simulate the system and observe the effect on opinion market shares. 
On Facebook, reputable media coverage reduces the prevalence of far-right opinions.
% decreases with more reputable (and less controversial) media coverage. 
On Twitter, both reputable and controversial media coverage suppress far-right opinions. 
However, for some opinions (like ``Mainstream media cannot be trusted"), reputable news backfires increasing far-right opinions share. 
% increase with reputable news coverage. 

\textbf{The main contributions of the work are as follows:}
\begin{enumerate}
    \item A novel two-tier model of the opinion ecosystem that allows studying opinion interactions through an economics-based cooperation-competition lens. 
    We introduce simulation and estimation algorithms and study the convergence with synthetic tests.
    \item A synthetic testbed to uncover interactions across sympathizers and opponents of far-right opinions and likely effects of positive interventions via media coverage.
    \item A curated dataset of Twitter and Facebook discussions on bushfires/climate change. 
\end{enumerate}

\textbf{Related Work.}
We focus the discussion of related work on models for cooperative-competitive interaction in a set of co-diffusing online items.
These models need to be both \textit{predictive} and \textit{interpretable} (usually generative models). 
We have observed a lack of recent research in this area, with few works dating after 2017.
%The Clash of the Contagions model \cite{Myers2012} frames interaction as the perturbation in the probability of \textit{infection} (i.e., adoption) of a given \textit{contagion} (i.e., online item), after exposure to a set of competing contagions. 
%The InterRate model \cite{Poux2021} approaches the task as a kernel estimation problem, inferring an \textit{interaction profile} that characterizes the temporal evolution of the interaction. 
Closely related to our proposal is the Correlated Cascades (CC) model \cite{Zarezade2017}, a variant of the multivariate Hawkes process to model product adoption across a set of competing products in a social network. 
It estimates the interaction parameter $\beta$, tuning the market cooperation or competition level. 
A limitation of CC is that all products share a single $\beta$ value. This simplifies existing asymmetric relationships and assumes that all brands either cooperate or compete. 
\omm addresses this issue by fully modeling these asymmetric relationships. 
Another closely related work is the Competing Products (CP) model \cite{Valera2015}, a multivariate Hawkes model for product adoption/use where the frequency of use is affected by the usage of other products. 
Limitations of the work are the absence of the assumption of limited attention and the possibility of negative intensities since competitive interactions are modeled as negative parameters.
\omm avoids the weaknesses of CP by using a multiplicative model, thereby avoiding negative intensities and defining opinion shares as fractions of the total attention volume. 
The SLANT model \cite{De2016} and the follow-up SLANT+ \citep{Kulkarni2017} extend the CP model to differentiate between a user's latent and expressed opinion and uses a similar Hawkes process to model the intensity. 
However, SLANT requires fine-grained network information for training, which is prohibitive considering that online platforms are becoming more stringent with fine-grained data access \citep{Venturini2019}. On the other hand, \omm requires only opinion counts for training.

\textbf{Ethics of Opinion Moderation and Broader Perspectives.}
\omm is intended to model interactions between opinions and be used as a testbed for evaluating positive interventions for opinion moderation.
As any tool, \omm is unaware of the intention of its user and, in theory, could be used by oppressive regimes to silence or manipulate the liberal opinions of their opponents \cite{Radsch2016}.
In addition, the fundamental value of freedom of speech for democratic societies implies that non-widely accepted opinions also have the right to be expressed.
% Granted, in the hands of a bad actor \omm may be used as a tool to limit freedom of speech, particularly in repressive governments aiming to censor liberal opinions.
The scientific literature studies this ethical conundrum in the context of Countering Violent Extremism (CVE) initiatives \cite{Betz2016,Radsch2016}.
When viewing \omm as an AI evaluation tool supporting CVE initiatives \cite{Ferguson2016}, these ethical issues can be mitigated using online CVE frameworks in liberal democracies \cite{Henschke2021}. 
We argue that the implementing body is responsible for \omm's ethical usage, and CVE regulations should be leveraged to mitigate malicious intent.

% \omm can be used as a testbed for evaluating interventions for opinion moderation.
% This brings \omm under the banner of Countering Violent Extremism (CVE) initiatives \cite{Betz2016,Radsch2016}.
% There have been significant debates concerning the ethics and justification of CVE, with critics flagging that an
% % There is controversy surrounding CVE initiatives given that 
% improper implementation has the potential to hamper freedom of speech and censor liberal opinions \cite{Radsch2016}. 
% % Ethical concerns and justifications for CVE have been debated, and 
% However, there exist frameworks for the implementation of online CVE in liberal democracies \cite{Henschke2021}. 
% Since \omm is an AI evaluation tool supporting CVE initiatives \cite{Ferguson2016}, the implementing body is responsible for regulating its ethical usage.
% % autonomy to enact morally grounded decisions.
\textbf{Causal Impact.} 
\omm measures the effect of media coverage on the opinion market shares using a generative model to disentangle endogenous and exogenous factors from observational data, similar to \cite{RizoiuXie2017,Fujita2018,Garetto2021}. 
Our model works on aggregate observational data (i.e., opinion counts), and it does not prove the causal impact of media coverage on individual opinion formation (i.e., behavior change).
We would require a pre-test/post-test control group design to achieve true causal links. 
Previous work \cite{King2017,Guess2021,Agovino2022} provides evidence of the interventional role of media coverage.
In \cref{subsec:what-if}, we explore this further in a what-if experiment to demonstrate how the level of media coverage affects opinion market shares.
% Given the evidence of the interventional role of media coverage in previous studies \cite{King2017,Guess2021,Agovino2022}, \hl{we consider its causal impact as a reasonable model assumption}\mar{I don't understand this part! Can you please rephrase or detail?}.

%engaged The media's role in mitigating extremism is facilitated by two different sectors with differing agendas: (1) the media development sector, whose primary goal is to promote a free, plural and inclusive media ecosystem, and (2) the security/military sector engaged in Countering Violent Extremism (CVE) initiatives, which pursue a strategic communication objective to counter propaganda and push forth alternative narratives \cite{Betz2016,Radsch2016}. CVE initiatives have been controversial given that their implementation in authoritarian regimes have the potential to censor liberal opinions \cite{Radsch2016}. Our model has utility in both sectors as an evaluation tool, which has been sparsely researched \cite{Ferguson2016}, but as with any CVE initiative ethical use lies within the implementing body. Continued collaboration and dialog between the media development and security/military sectors should be promoted to ensure checks and balances are in place to ensure freedom of speech \cite{Radsch2016}.

\section{Preliminaries}
We introduce two classes of models that form the foundation of our approach: 
(1) the discrete-time Hawkes process \cite{Browning2021}, a model of event counts that display self-exciting behavior, and 
(2) the market share attraction model \cite{Cooper1993}, a marketing model that uncovers the latent competitive structure of brands and interaction with marketing instruments.

\subsection{Discrete-time Hawkes Process}
The discrete-time Hawkes Process (DTHP) \cite{Browning2021} is the discrete-time analogue of the popular self-exciting Hawkes process \cite{Hawkes1971}, where instead of modeling the occurrence of events given by $t \in \mathbb{R}^+$, we model the event count $N(t)$ on $[t-1, t)$ for $t \in \mathbb{N}$. 

The DTHP is characterized by the conditional intensity function $\lambda(t)$, defined as the expected number of events that occur at time $t$, conditioned on the history $H_{t-1} = \{N(s)| s < t\}$.
For a DTHP, $\lambda(t)$ is given by
\begin{equation}
    \lambda(t) = \mathbb{E}[N(t)|H_{t-1}] = \mu + \sum_{s<t} \alpha \cdot f(t-s) \cdot N(s) , \label{eq:dthp_lambda}
\end{equation}
where $\mu$ is the baseline count of events, $\alpha$ determines the level of self-excitation and is the expected number of events produced by a single event and $f(t)$ is the triggering kernel, which controls the influence of the past events on the present. We specify $f(t)$ with the geometric probability mass function $f(t) = \theta (1-\theta)^{t-1}, t \in \mathbb{N}$, the discrete-time analogue of the exponential distribution \cite{Browning2021}. Given $\lambda(t)$, model specification is completed by specifying a probability mass function for the count $N(t)$. Following \cite{Browning2021}, we set  $N(t) \sim \text{Poi}(\lambda(t))$.

%% NONESSENTIAL
% Given a dataset of event counts $\{n_t | t \in \{1, \ldots, T\}\}$, with $n_t$ counting the number of events on $[t-1,t)$, one can determine the best-fitting DTHP parameters $\{\mu, \alpha, \theta\}$ by maximizing the log-likelihood function:

% \begin{multline}
%     \mathcal{L}(\mu, \alpha, \theta | n_1 \ldots n_T) \\
%     \propto \sum_{t=1}^T \Bigg[ n_t \log \left(\mu + \alpha \sum_{s < t} n_s \theta (1-\theta)^{t-s-1} \right) \\
%     - \left( \mu + \alpha \sum_{s < t} n_s \theta (1 - \theta)^{t-s-1} \right) \Bigg].
% \end{multline}

\subsection{Market Share Attraction Model}

In marketing literature, \textit{market share attraction models} (MSAMs) \cite{Cooper1993} model the competitive structure of a set of $M$ brands in a product category,  predict their market shares, and evaluate how a set of marketing instruments affects resulting market shares. 
% We cast online opinions as brands competing for limited attention and interventions as marketing instruments.
% 
MSAMs assume that the market share $s_i$ of brand $i \in \{1 \ldots M\}$ is proportional to consumers' attraction $\mathcal{A}_i$ towards brand $i$:
\begin{equation}
    s_i = \frac{\mathcal{A}_i}{\sum_{j=1}^M \mathcal{A}_j} \in [0,1]. \label{eq:siai}
\end{equation}
% 
% Brand $i$'s attraction 
$\mathcal{A}_i$ is typically modeled as a parametric function of a set of $K$ marketing instruments $\{X_{ki}\}_{k=1}^K \in \mathbb{R}^K$, where $X_{ki}$ gives the value of the $k^{th}$ marketing instrument for brand $i$. 
% Typical examples of marketing instruments are the price of the product, ad spend, and distribution efforts.
% 
% A common specification of the attraction model is the multinomial logit (MNL) model given by 
We typically specify $\mathcal{A}_i$ as
\begin{equation}
    \mathcal{A}_i = \exp\left(\beta_i + \sum_{k=1}^K \sum_{j=1}^M \gamma_{kij} X_{kj} \right), \label{eq:mnl}
\end{equation}
where $\beta_i$ measures the inherent attraction of brand $i$ and $\gamma_{kij} \in \mathbb{R}$ measures the effect of the value of the $k^{th}$ marketing instrument for brand $j$ on brand $i$'s attraction. Whether $\gamma_{kij}$ is positive (negative) is indicative of the excitatory (inhibiting) relationship from brand $j$ to brand $i$ through marketing instrument $X_{kj}$.

MSAMs are interpreted via the model elasticity $e(s_i, X_{kj})$, the ratio of the percent change in the market share $s_i$ given a percent change in the value of the $k^{th}$ marketing instrument for brand $j$. 
For example, an elasticity of $e(s_i, X_{kj}) = 0.1$ means that a $1\%$ increase in $X_{kj}$ corresponds to a $0.1\%$ increase in $s_i$. 
That is,
\begin{equation}
    e(s_i, X_{kj}) = \frac{\partial s_i / s_i}{\partial X_{kj} / X_{kj}} = \frac{\partial s_i}{\partial X_{kj}} \cdot \frac{X_{kj}}{s_i}.
    \label{eq:elasticity}
\end{equation}

The elasticity $e(s_i, X_{kj})$ captures the overall effect of brand $j$'s marketing instrument $X_{kj}$ on brand $i$'s market share $s_i$: both the \textit{direct effect} of $X_{kj}$ on $s_i$, controlled by $\gamma_{kij}$, and the \textit{indirect effect} of $X_{kj}$ on $s_i$ through its effect on the attraction of other brands $\{j \neq i\}$. 
% Given a marketing instrument $X_{k}$, the elasticities $\{e(s_i, X_{kj})\}_{ij}$ form a matrix summarizing the cooperative-competitive interactions across the set of brands.

\section{The \omm Model}
\label{section:model}

In this section, we develop a two-tier model of the opinion ecosystem. The first tier models the total size of the opinion attention market on multiple online platforms. The second tier models the market share of opinions on each platform. Next, we 
%present an algorithm to sample opinion volumes from a fitted model and 
introduce a scheme for parameter estimation.

% \subsection{Formulation}
\label{subsection:formulation}

%Consider a set of $M$ opinions that diffuse across $P$ online platforms over observation days. Let $N^p_i(t)$ be the number of posts o n platform $p \in \{1, \ldots P\}$ expressing opinion $m \in \{1, \ldots, M\}$ on day $t \in \mathbb{N}$. We model $N^p_i(t)$ as a random variable with conditional intensity $\lambda^p_i(t)$, i.e.
% \begin{equation}
%     N^p_i(t) \sim \text{Poi}(\lambda^p_i(t)).
% \end{equation}
% Furthermore, we define $N^p(t)$ as the total number of opinionated posts on platform $p$ on day $t$. Let  $\lambda^p(t)$ be the conditional intensity of $N^p(t)$. Then,
% \begin{equation}
    % N^p(t) = \sum_{i=1}^M N^p_i(t) \sim \text{Poi}(\lambda^p(t)).
% \end{equation}

% We assume that $\lambda^p_i(t)$ splits as a product of two processes: (1) the expected opinion count on platform $p$ on day $t$, and (2) the market share of opinion $i$ on platform $p$ on day $t$. That is,\\
\omm consists of two tiers; the first tier, which we call the \textit{opinion volume model}, tracks the size of the opinion attention market, while the second tier, the \textit{opinion share model}, tracks the market shares of the different opinions. \cref{tab1:tabnotations} summarises the notation for important variables in the \omm. The full table is available in the online appendix \citep{appendix}.

\textbf{Opinion Volume Model.} Suppose our opinion ecosystem consists of $P$ social media platforms. % on which users interact and post their opinions. 
The opinion volume model tracks the attention volume, i.e. the number of opinionated posts $N^p(t)$, on each platform $p \in \{1, \ldots P\}$ and time $t \in \mathbb{N}$. 
We model $\{N^p(t)\}_p$ as a $P-$dimensional DTHP (defined analogous to the multivariate Hawkes process \cite{Hawkes1971}) with conditional intensity $\{\lambda^p(t)\}_p$,
\begin{equation}
    \lambda^p(t) = \mu^p \cdot S(t) + \sum_{q=1}^P \sum_{s<t} \alpha^{pq} \cdot f(t-s) \cdot N^q(s).
    \label{eq:opinionvolumemodel}
\end{equation}

In contrast to \cref{eq:dthp_lambda}, we use a time-varying exogenous signal $S(t)$, which accounts for the baseline volume of events of exogenous origin. The signal $S(t)$ accounts for natural tendencies and events (i.e., epidemics, elections) and typically cannot be controlled.
We introduce a scaling term $\mu^{p}$ for each platform $p$ such that $\mu^p \cdot S(t)$ represents the exogenous opinion count for platform $p$ on time $t$.

Since online platforms are not siloed and have significant user overlap, we allow the $P$ platforms to interact via intra- and inter-platform excitation. The parameter $\alpha^{pq} > 0$ sets the level of intra-platform (for $p=q$) and inter-platform (for $p \neq q$) excitation. Lastly, we set $N^p(t) \sim \text{Poi}(\lambda^p(t))$.

\textbf{Opinion Share Model.} With the attention volumes for each platform $p$ estimated in the first tier, the second tier models the market share $s_i^p(t)$, calculated as the fraction of opinionated posts on platform $p$ conveying opinion $i$. Given the limited attention market size, opinions compete for attention within each platform. 

Suppose that there are $M$ different opinion types. We set $N^p_i(t)$ to be the number of opinionated posts conveying opinion $i$ on platform $p$ on time $t$, and $\lambda^p_i(t)$ to be its conditional intensity. Using the notion of limited attention \cite{Zarezade2017}, we relate $\lambda^p_i(t)$ to $\lambda^p(t)$ in \cref{eq:opinionvolumemodel} by introducing the market share $s^p_i(t) \in [0,1]$ as the fraction of opinion $i$ posts on platform $p$. That is,
\begin{equation}
    \lambda^p_i(t) = \lambda^p(t) \cdot s_i^p(t), \label{eq:lambbda_pi_omm}
\end{equation}
and $\sum_{i=1}^M s_i^p(t) = 1$.
% This allows us to split $N^p(t)$ as a set of $M$ variables $\{N^p_i(t)\}_i$, i.e.
% \begin{equation}
%     N^p_i(t) = \sum_{i=1}^M N^p_i(t) \sim \text{Poi}(\lambda^p_i(t)).
% \end{equation}

Similar to \cref{eq:siai}, we model $s^p_i(t)$ with attraction $\mathcal{A}^p_i(t)$,
\begin{equation}
    s_i^p(t) = \frac{\mathcal{A}^p_i(t)}{\sum_{j=1}^M \mathcal{A}^p_j(t)}. \label{eq:attraction_fraction}
\end{equation}
Leveraging the MNL form in \cref{eq:mnl}, we define attraction

% which in our model is dependent on \textit{endogenous} opinion dynamics and \textit{exogenous} factors. Given $\mathcal{A}^p_i(t)$, we are able to calculate $s_i^p(t)$, the market share of opinion $i$ on platform $p$ on day $t$ as follows:

\begin{equation}
    % \mathcal{A}_i^p(t) = \exp \left(\underbrace{b_i^p}_{\text{base tendency}} + 
    \mathcal{A}_i^p(t) = \exp \mathcal{T}_i^p(t),
    \label{eq:attraction_tendency}
\end{equation}
where $\mathcal{T}_i^p(t)$ consists of two parts, accounting for \textit{interventions} and \textit{endogenous} dynamics, and is described in detail below,
\begin{equation}
    % \mathcal{A}_i^p(t) = \exp \left(\underbrace{b_i^p}_{\text{base tendency}} + 
    \mathcal{T}_i^p(t) =
    \underbrace{\sum_{k=1}^K  \gamma_{ik}^{p}  \cdot \bar{X}_k(t)}_{\text{interventions}} + \underbrace{\sum_{q=1}^P \sum_{j=1}^M \beta_{ij}^{pq} \cdot \lambda^q(t|j)}_{\text{endogenous}},
    \label{eq:opinionsharemodel_tendency}
\end{equation}
\begin{equation}
    \bar{X}_k(t) = \sum_{s < t} f(t-s) \cdot X_{k}(s), \quad \text{and} \nonumber
\end{equation}
\begin{equation}
    \lambda^p(t|j) = \mu^p_j \cdot S(t) + \sum_{q=1}^P \sum_{s<t} \alpha^{pq} \cdot f(t-s) \cdot N^q_j(s),
    \label{eq:conditional_j}
\end{equation}
where $\mu^p = \sum_{j=1}^M \mu^p_j$.

In the first term of \cref{eq:opinionsharemodel_tendency}, we introduce a set of $K$ positive interventions $\left\{X_{k}(t)\right\}_k$ that modify the opinion market shares in the opinion ecosystem. The interventions $\left\{X_{k}(t)\right\}_k$ have a different to $S(t)$ in \cref{eq:opinionvolumemodel}, as the latter modifies the attention market size. Parameter $\gamma_{ik}^{p} \in \mathbb{R}$ measures the direct effect of the $k^{th}$ intervention on the market share of opinion $i$ on platform $p$. If $\gamma_{ik}^{p}$ is positive (negative), then $X_{k}(t)$ reinforces (inhibits) opinion $i$ on platform $p$.

In the second term of \cref{eq:opinionsharemodel_tendency} we model the contribution of endogenous dynamics on the attraction of opinion $i$. To represent the prevalence of opinion $j$ on platform $q$, we make use of the conditional intensity $\lambda^p(t|j)$ in \cref{eq:conditional_j}, which models the dynamics of opinion $j$ independent of other opinions. Parameter $\beta_{ij}^{pq} \in \mathbb{R}$ captures the direct effect that opinion $j$ on platform $q$ has on the market share of opinion $i$ on platform $p$. 
Similar to $\gamma_{ik}^{p}$, we allow $\beta_{ij}^{pq}$ to be positive (negative), representing a reinforcing (inhibiting) relationship from opinion $j$ to $i$ on platform $q$ and $p$, respectively.

\begin{table}[t!]
    \begin{tabularx}{\columnwidth}{p{0.14\columnwidth} p{0.81\columnwidth}}
        \toprule
        Notation & Interpretation \\
        \midrule
        $P$  & number of social media platforms \\
        $M$ & number of opinion types  \\
        $K$  & number of positive interventions  \\
        $T$  & terminal time  \\
        \bottomrule
        \toprule
        Variables & \\
        \midrule
        $S(t)$  & input signal, volume of exogenous events  \\
        $X_{k}(t)$  & input signal, $k^{th}$ positive intervention  \\
        $s^p_i(t)$  & market share of opinion $i$ on platform $p$ at time $t$ \\
        $\lambda^p_i(t)$ & conditional intensity of opinion $i$ \\
        $N^p_i(t)$  & \#posts with opinion $i$ on platform $p$ at time $t$ \\
        $e(s^p_i(t), \cdot)$ & opinion share model elasticity \\
        \bottomrule
        \toprule
        Data & \\
        \midrule
        $n^p_t / n^p_{i,t}$  & \#posts on platform $p$ at time $t$ / with opinion $i$  \\
        $s^p_{i,t}$  & fraction of posts on platf. $p$ with opin. $i$ at time $t$  \\
        \bottomrule
        \toprule
        Parameters & \\
        \midrule
        $\mu_j^p$  & exogenous scaling term for opin. $j$ on platf. $p$ \\
        $\alpha^{pq}$  & excitation parameter for intra-platform ($p=q$) and inter-platform (for $p \neq q$) dynamics  \\
        $\theta$  & memory parameter, describing how fast an event is forgotten, $\theta \in [0,1]$ \\
        $\gamma^p_{ik}$  & direct effect of the $k^{th}$ intervention on share of opinion $i$ on platform $p$  \\
        $\beta^{pq}_{ij}$  & direct effect that opinion $j$ on platform $q$ has on share of opinion $i$ on platform $p$.  \\
        \bottomrule
    \end{tabularx}
    \caption{Summary of important quantities and notations.}
    \label{tab1:tabnotations}
\end{table}

\textbf{Estimation.}
\label{subsection:estimation}
Over the observation period $t \in \{1, \ldots, T\}$, assume that we observe the exogenous signal $S(t)$, the $K$ interventions $\{X_k(t)\}_k$, and the number $n^p_{i,t}$ of posts conveying opinion $i$ on platform $p$ for each $i$ and $p$. Our goal is to estimate the parameter set $\boldsymbol{\Theta} = \{\mu_j^p, \alpha^{pq}, \theta, \gamma^p_{ik}, \beta^{pq}_{ij}\}$.

The structure of our two-tier model allows us to cast parameter estimation as a two-tier optimization problem. Let $\boldsymbol{\Theta}_1 = \{\mu^p, \alpha^{pq}, \theta \}$. The key observation here is that the first-tier parameter set $\boldsymbol{\Theta}_1$ can be estimated using only the opinion volume model in \cref{eq:opinionvolumemodel}, independent of the opinion share model in \cref{eq:opinionsharemodel_tendency}. By optimizing the likelihood $\mathcal{L}_1(\boldsymbol{\Theta}_1| \{n^p_t\}_{p,t})$ of the platform-level volumes $\{n^p_{t}\}_{p,t}$, we can obtain an estimate $\hat{\boldsymbol{\Theta}}_1$ of $\boldsymbol{\Theta}_1$. 

The second-tier parameter set $\boldsymbol{\Theta}_2 = \{\mu^p_j, \gamma^p_{ik}, \beta^{pq}_{ij}\}$ can be obtained by optimizing the likelihood $\mathcal{L}_2(\boldsymbol{\Theta}_2 | \hat{\boldsymbol{\Theta}}_1, \{n^p_{i,t}\}_{i,p,t})$ of the opinion volumes $\{n^p_{i,t}\}_{i,p,t}$, conditioned on our estimate of the first-tier parameters $\hat{\boldsymbol{\Theta}}_1$. %We summarize parameter estimation below. 
Our full estimated parameter set is given by $\hat{\boldsymbol{\Theta}} = \hat{\boldsymbol{\Theta}}_1 \cup \hat{\boldsymbol{\Theta}}_2$. 
The technical details of the estimation and the derivation of the likelihoods $\mathcal{L}_1(\cdot)$ and $\mathcal{L}_2(\cdot)$ and gradients $\partial_{\boldsymbol{\Theta}_1}\mathcal{L}_1(\cdot)$ and $\partial_{\boldsymbol{\Theta}_2}\mathcal{L}_2(\cdot)$ are available in the online appendix \cite{appendix}.

\begin{figure*}
    \newcommand\myheight{0.20}
    \centering
    \subfigure[]{
        \includegraphics[height=\myheight\textheight]{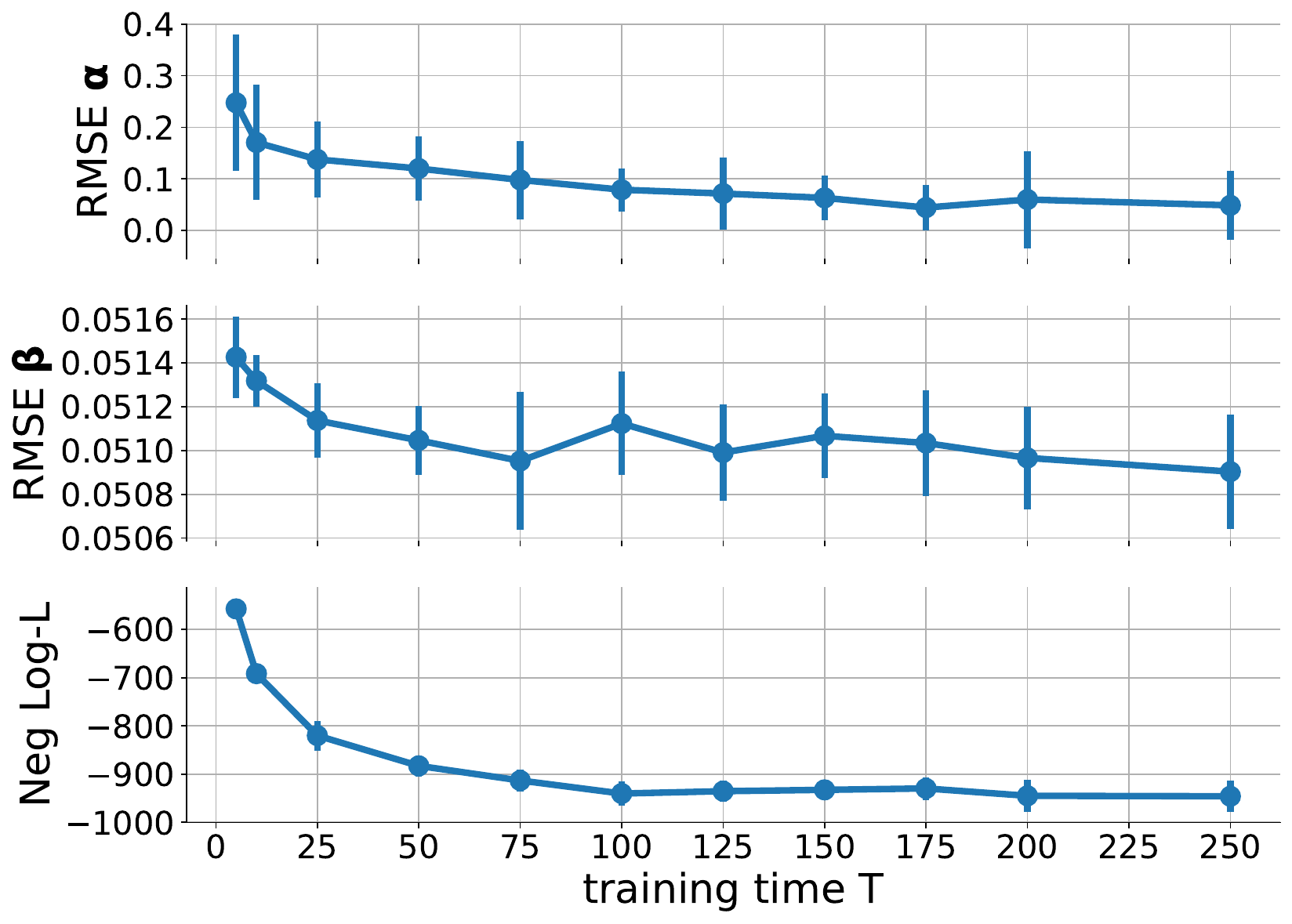}
        \label{fig:synthetic-a}} 
    \subfigure[]{
        \includegraphics[height=\myheight\textheight]{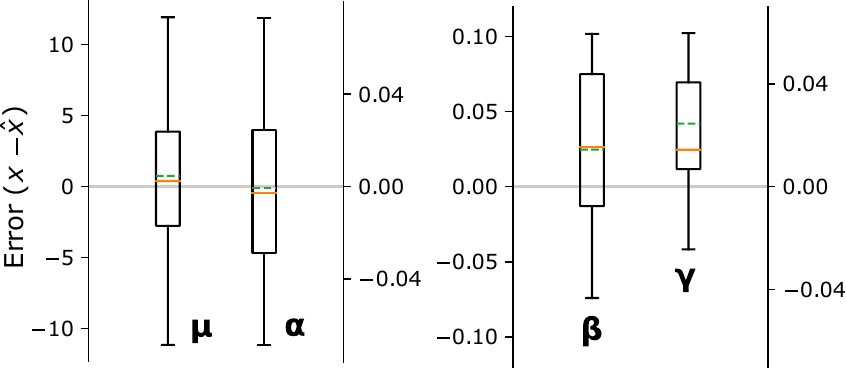}
        \label{fig:synthetic-perc}
    } 
    \caption{Parameter recovery results on synthetic data. In (a), we show the convergence of the RMSE of the $\boldsymbol{\alpha}$ and $\boldsymbol{\beta}$ estimates and the negative log-likelihood as we increase the training time $T$. In (b), we show the difference between our estimates for $\{\boldsymbol{\mu}, \boldsymbol{\alpha}, \boldsymbol{\beta}, \boldsymbol{\gamma}\}$ and the true values. Dashed \textcolor{green}{green} lines and \textcolor{orange}{orange} lines are the mean and median values, respectively.}
    \label{fig:synthetic}
\end{figure*}

\textbf{Simulation.}
Suppose we are given the opinion volume $n^p_{i,0}$ at time $t=0$ for each platform $p$ and opinion $i$, such that $n^{p}_{t} = \sum_i n^p_{i,t}$. 
A sample of $n^{p}_{i,t}$ from \omm can be obtained by calculating the conditional intensity $\lambda^p_i(t)$ using \cref{eq:lambbda_pi_omm}, and then sampling $n^p_{i,t}$ from Poi ($\lambda^p_i(t)$). We obtain $\{n^p_{i,t}\}_{i,p,t}$ by repeating these steps over $\{1, \ldots, T\}$.% and each platform $p$ and opinion $i$.

% \begin{enumerate}
%     \item Compute $\lambda^p_i(t) = \lambda^p(t | \cup_{q,s<t}\{ n^q_s\}) \cdot s^p_i(t | \cup_{q,j, s<t} \{ n^q_{j,s} \})$.
%     \item Draw a sample $n^p_{i,t} \sim \text{Poi}(\lambda^p_i(t))$.
% \end{enumerate}

\textbf{Numerical Considerations.} 
To improve model fit in our real-world case studies, we implement three augmentations to the model and estimation method, outlined below and fully detailed in the online appendix \cite{appendix}.
First, we modify the attraction $\mathcal{A}_i^p(t)$ in \cref{eq:attraction_fraction} to prevent numerical overflow/underflow. 
Second, we add a regularization term in the second-tier optimization problem in \cref{subsection:estimation} to impose structural constraints on $\{\hat{\gamma}^p_{ik}\}$ and improve estimation. 
Third, we apply log-scaling on $\lambda^q(t|j)$ and standardize both $\lambda^q(t|j)$ and $\bar{X}_k(s)$ in \cref{eq:opinionsharemodel_tendency} to solve scaling issues.

\textbf{Stability Assumption.} We implicitly assume that the opinion attention market is stable over the timeframe of the analysis, in the sense that the parameters $\boldsymbol{\Theta}$ governing the behavior of the process stay constant within the timeframe. In situations where this assumption is not expected to hold (e.g. extreme events) and parameters change within the timeframe, a change-point model extension \citep{Browning2021} of the \omm is necessitated.

\section{Learning with Synthetic Data}
In this section, we consider the parameter estimation task with synthetic data. 
First, we discuss our experimental setup and the synthetic dataset. 
Next, we show that parameter recovery error decreases and stabilizes as we increase the training time $T$ and the number of samples $n_{samples}$. 

\textbf{Experimental Setup.} 
We set $P = M = K = 2$. 
We set $[\mu^1_1, \mu^1_2, \mu^2_1, \mu^2_2] = [15,5,5,20]$, and $\theta = 0.5$
% $\mu^1_1 = 15$, $\mu^1_2 = 5$, $\mu^2_1 = 5$, $\mu^2_2 = 20$, $\theta = 0.5$, 
and draw $\alpha^{pq} \sim \text{Unif}(0,0.5)$, $\beta^{pq}_{ij} \sim \text{Unif}(0,0.1)$ and $\gamma^{p}_{ik} \sim \text{Unif}(0,0.1)$. 
The exogenous signals are $S(t)=1$, $X_1(t) = 5 \sin (0.1 x) + 5$, and  $X_2(t) = 10 \sin (0.05 x + 1.25)+10$.

We construct our synthetic dataset using the simulation algorithm in \cref{subsection:estimation} to get 400 samples of opinion volumes $\left\{n_{i,t}^p \right\}_{i,p,t}$ for $t \in \{1, \ldots, T=300\}$.
We implement joint fitting \cite{Rizoiu2022}: we partition the 400 samples into 20 groups of $n_{samples} = 20$ samples each. The likelihoods $\mathcal{L}_1(\boldsymbol{\Theta}_1)$ and $\mathcal{L}_2(\boldsymbol{\Theta}_2 | \boldsymbol{\Theta}_1)$ of each group are maximised to get an estimate $\hat{\boldsymbol{\Theta}}$, yielding 20 sets of parameter estimates.

\textbf{Model Evaluation.} To study the convergence of our learning algorithm, we compute the root mean-squared error (RMSE) of our estimated $\hat{\boldsymbol{\Theta}} = \{\hat{\mu}_j^p, \hat{\alpha}^{pq}, \hat{\theta}, \hat{\gamma}^p_{ik}, \hat{\beta}^{pq}_{ij}$\} with respect to the true $\boldsymbol{\Theta}$, following \cite{Valera2015}. 
We report the average RMSE per parameter type, where the average is taken over the components of the matrix or tensor corresponding to the parameter type.

In \cref{fig:synthetic-a}, we see that training on a longer timeframe leads to lower RMSE for $\hat{\alpha}^{pq}$ and $\hat{\beta}^{pq}_{ij}$ and better model fit measured by the likelihood $\mathcal{L}_2$. 
Results for $\hat{\mu}_j^p$, $\hat{\theta}$ and $\hat{\gamma}^p_{ik}$, and on varying $n_{samples}$ are in the online appendix \cite{appendix}.

In \cref{fig:synthetic-perc}, we plot the difference distribution between our estimates and the true values. 
We recover first-tier parameters $\{\hat{\mu}_j^p, \hat{\alpha}^{pq}\}$ well, as evidenced by our mean estimates coinciding with the true values. 
%For the second-tier parameters $\{\hat{\gamma}^p_{ik}, \hat{\beta}^{pq}_{ij}\}$ we observe slight overestimation.
We observe a slight overestimation of $\{\hat{\gamma}^p_{ik}, \hat{\beta}^{pq}_{ij}\}$, given the nonconvexity of $\mathcal{L}_2$ and the high dimensionality of the second-tier parameter set.

\section{Real-World Datasets}
\label{subsection:real}

This section introduces two real-world datasets we have curated to evaluate the \omm.
% In this section, we evaluate \omm on a dataset of discussions annotated with moderate and far-right opinions.
% First, we introduce the \textit{Bushfire Opinions dataset}, the exogenous signal $S(t)$ and positive interventions $\{X_k(t)\}$.
% Second, we evaluate \omm's performance in predicting volumes and market shares of opinions.
% Lastly, we interpret the model elasticities to understand opinion relations and perform what-if scenarios to uncover the effect of the intervention.

\subsection{Bushfire Opinions dataset}
\label{subsec:dataset}
We construct the \textit{Bushfire Opinions dataset}, containing 90 days of Twitter and Facebook discussions about bushfires and climate change between November 1, 2019 to January 29, 2020.
The Facebook postings are a subset of the \textit{SocialSense} dataset \cite{Kong2021};
we select posts and comments about bushfires and climate change (\textit{SocialSense} also contains discussions around COVID-19).
These were collected using CrowdTangle\footnote{
https://www.crowdtangle.com/} by crawling public far-right Australian Facebook groups, identified via a digital ethnographic study (see \cite{Kong2021} and the online appendix~\cite{appendix} for details).
We build the Twitter discussions using the Twitter Academic v2 API;
we collect tweets emitted between November 1, 2019 to January 29, 2020 that mention bushfire keywords such as \textit{bushfire}, \textit{arson}, \textit{australiaburns} (see the full list in the online appendix~\cite{appendix}).
We use the AWS Location Service\footnote{https://aws.amazon.com/location/} to geocode users based on their free-text location and description fields and filter only for tweets from Australian users.

Our focus on the 2019-2020 Australian bushfires is motivated by the availability of human-annotated topics, opinions \cite{Kong2021} and stance classifiers \cite{Ram2022} trained on the same topic and timeframe. 
We use these classifiers to filter and label our dataset.

\textbf{Moderate and Far-Right Opinion Labeling.}
To filter and label relevant Facebook and Twitter postings, we use the textual topic and opinion classifiers developed by \citet{Kong2021}, with a reported $93\%$ accuracy in classifying Facebook and Twitter posts on bushfires and climate change.
We select the following most prevalent six opinions, covering $95\%$ of Twitter and $81\%$ of Facebook postings:
\begin{enumerate}\addtocounter{enumi}{-1}
    \item Greens policies are the cause of the Australian bushfires.
    \item Mainstream media cannot be trusted.
    \item Climate change crisis is not real / is a UN hoax.
    \item Australian bushfires and climate change are not related.
    \item Australian bushfires were caused by random arsonists.
    \item Bushfires are a normal summer occurrence in Australia.
\end{enumerate}
Furthermore, we deploy the far-right stance detector introduced by \citet{Ram2022} -- which leverages a textual homophily measurement to quantify the similarity between Twitter users and known far-right activists. 
On the \textit{Bushfire Opinions Twitter dataset}, the stance detector achieves a 5-fold CV AUC ROC score of $0.889$.
An opinion is labeled as \emph{far-right} if the posting agrees with the opinion (denoted as +), or \emph{moderate} if the posting disagrees with the opinion (-).
We represent our opinion set as $\{(i-,i+) | i \in \{0, \ldots, 5\}\}$.
In summary, we consider $P=2$ platforms with $74,461$ tweets and $7,974$ Facebook postings labeled with $M=12$ stanced opinions.
We aggregate posting volumes by the hour, resulting in $T=2,160$ time points over 90 days from Nov 1, 2019, to Jan 29, 2020.

\begin{figure*}[!tbp]
    \newcommand\myheight{0.33}
    \centering
    \subfigure[]{
        \includegraphics[height=\myheight\textheight]{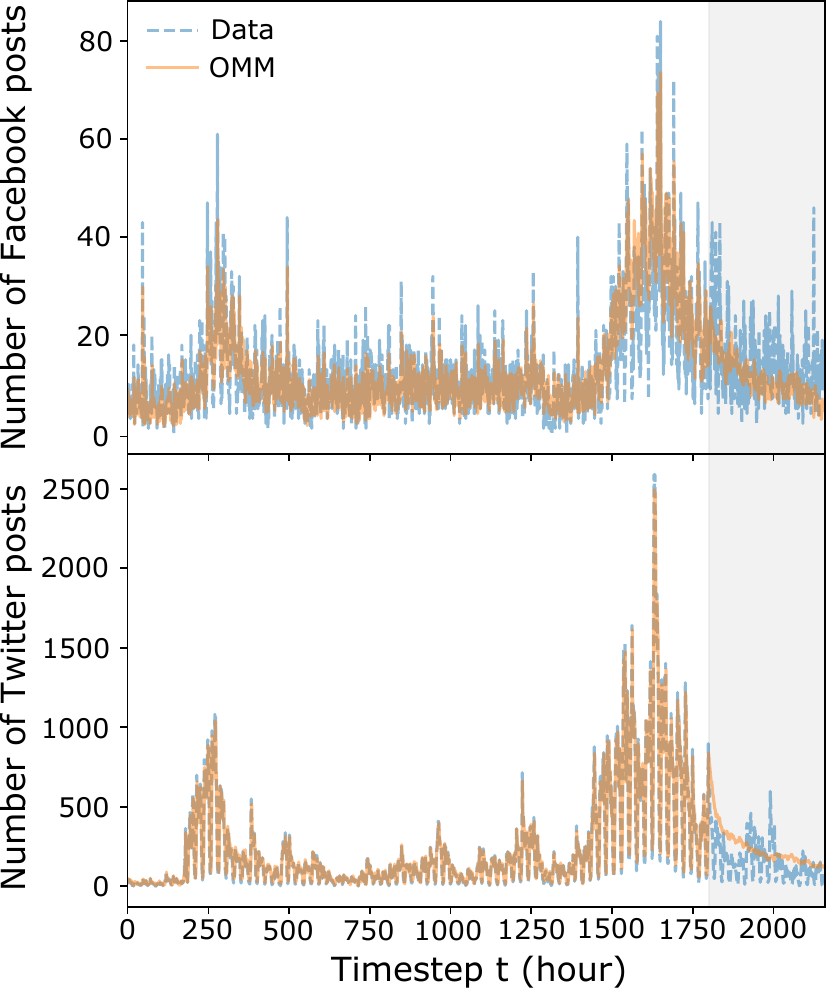}
        \label{fig:level1}
    } 
    \subfigure[]{
        \includegraphics[height=\myheight\textheight]{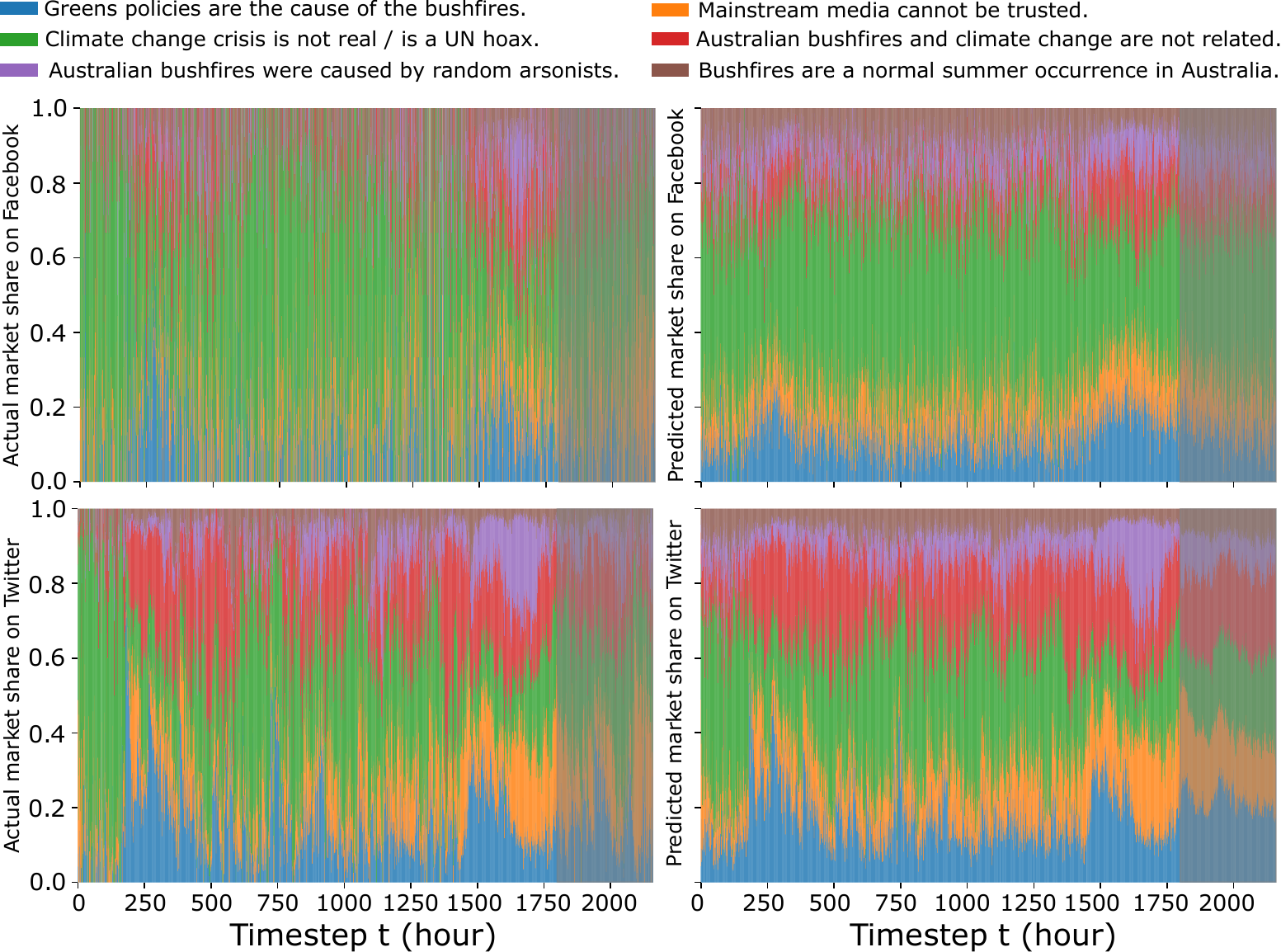}
        \label{fig:level2}
    } 
    % \subfigure[]{
    %     \includegraphics[height=\myheight\textheight]{plots/improved_fig3/kl_v2.pdf}
    %     \label{fig:kl}
    % } 
    \caption{
        Fitting and predicting with \omm on the Bushfire Opinions dataset. 
        We train \omm on the first 1800 timesteps and predict on timesteps 1801 to 2160 (shaded area). 
        We show results for Facebook (top row) and Twitter (bottom row).
        (a) Actual (dashed \textcolor{blue}{blue} lines) vs. fitted/predicted  (\textcolor{orange}{orange} lines) volumes; 
        (b) Actual (left panels) and fitted during training and predicted during testing (right panels) opinion market shares on Facebook and Twitter.
        We aggregate the far-right and moderate opinions. 
    }
    % (c) KL-divergence on the evaluation set for the CC and CP baselines and \omm with and without media coverage $X(t)$.}
    \label{fig:performance}
\end{figure*}

\textbf{Exogenous Signal S and Intervention X.} 
%% MAR: describe one, then the other, don't mix them
The exogenous signal $S(t)$ (\cref{eq:opinionvolumemodel}) modulates the total size of the attention market in the first tier of \omm.
We use the 5-day rolling average of the Google Trends query \textit{bushfire+climate change} in Australia, normalized to a max value of 1. 
Google Trends captures the baseline interest on topics \cite{Sheshadri2019} and is a proxy for offline events (ex. actual bushfires and government measures) \cite{Milinovich2014}.

The interventions $\{X_k(t)\}$ modulate the market share of far-right and moderate opinions.
Our interventions consist of two sources of news coverage: 
reputable ($R$) mainstream Australian publishers (e.g., The Sydney Morning Herald, Canberra Times, Crikey) and controversial ($C$) international publishers (e.g., Sputnik News, Breitbart, Red State). 
For each opinion $i \in \{0,\ldots,5\}$, we consider a pair of interventions $(R_i(t), C_i(t))$, consisting of reputable and controversial daily news volumes discussing opinion $i$. 
We assemble the intervention set $\{X_k(t)\}$ ($K=12$) so that the first six interventions correspond to $\{R_0(t), \ldots, R_5(t)\}$ while the last six correspond to $\{C_0(t), \ldots, C_5(t)\}$. 

We sourced reputable Australian news publishers from the Reputable News Index (RNIX) \cite{Kong2020}.
We query Factiva \cite{Factiva2009} to obtain the daily news volume of these outlets for each of the six opinions using a keyword search. 
We similarly obtain the news volumes from controversial international publishers from NELA-GT-2019 \cite{Gruppi2020} using a keyword search.
We subtract the Google Trends signal from the news volumes for each intervention. 
We compute the standardized form of $X_{k}(t)$ as 
$\hat{X}_{k}(t) = \text{news}_{k}(t) - \frac{\max_t \text{news}_{k}(t)}{\max_t S(t)} S(t)$. 
For brevity, in the bushfire case study, we denote $\hat{X}_{k}(t)$ as $X_{k}(t)$ (i.e., always in standardized form).
Standardization allows $X_{k}(t)$ to be interpreted as the extent to which reputable or controversial media over- or under-reports relative to the public's attention.

\subsection{VEVO 2017 Top 10 dataset}
We assemble the \textit{VEVO 2017 Top 10} dataset by aligning artist-level time series of YouTube views and Twitter post counts ($P=2$) for the top $M=10$ VEVO-affiliated artists over $T=100$ days from Jan 2, 2017 to Apr 11, 2017.
%The artists we consider are Justin Bieber, Katy Perry, Taylor Swift, Rihanna, Eminem, Maroon 5, Calvin Harris, Ariana Grande, One Direction, and Beyoncé.

The YouTube time series are obtained from the \textit{VEVO Music Graph dataset} \cite{Wu2019}, containing daily view counts for music videos posted by verified VEVO artists in six English-speaking countries (USA, UK, Canada, Australia, New Zealand, and Ireland). 
We combine the view counts for all music videos that belong to a given artist to obtain artist-level YouTube view time series.
For Twitter, we leverage the Twitter API to get daily counts of posts with text containing an input query. 
We obtain the artist-level Twitter post time series using the artist's name as the input query. 

Unlike the single exogenous signal $S(t)$ in the Bushfire Opinions dataset, we use a different exogenous signal $S_i(t)$ for each artist $i$ -- the Google Trends for each artist $i$. 
Using the set $\{S_i(t)\}$ instead of a single $S(t)$ requires several small changes to \cref{eq:opinionvolumemodel}, \cref{eq:conditional_j}, and the model gradients.
We fully detail these changes in the online appendix \cite{appendix}. 
We do not consider any interventions $\{X_k(t)\}$ as we seek to uncover endogenous interactions across artists.

\section{Predictive Evaluation}
\label{subsec:predictive-XP}
This section evaluates the \omm's predictive capabilities on two real-world datasets. 
We introduce our prediction task, evaluation metrics and baselines, then present the results.

\textbf{Model Setup.} 
We use a temporal holdout strategy similar to prior literature \cite{Rizoiu2017,Rizoiu2018,Kong2020}: 
we fit \omm on $\mathcal{T}_{obs}$ and evaluate performance on $\mathcal{T}_{pred}$. 
Backtesting is another viable alternate evaluation approach; however, it is significantly more computationally intensive, and we prefer the temporal holdout.
For the bushfire case study, $\mathcal{T}_{obs} = \{1, \ldots, 1800\}$ where time is in hours (i.e., days 1-75 of our period of interest) and $\mathcal{T}_{pred} = \{1801, \ldots, 2160\}$ (i.e., days 76-90). 
For the VEVO case study, $\mathcal{T}_{obs} = \{1, \ldots, 75\}$ and $\mathcal{T}_{pred} = \{76, \ldots, 100\}$. 

We consider two tasks: (1) opinion volume prediction and (2) opinion share prediction. 
For the first task, we predict the total volume of opinionated posts on the $P$ platforms during the evaluation period. 
We measure performance using the platform-averaged symmetric mean absolute percentage error (SMAPE) of predicted volumes $\{\bar{n}^p_t | t \in\ \mathcal{T}_{pred}\}$ on platform $p$ relative to the actual volumes $\{n^p_t | t \in\ \mathcal{T}_{pred}\}$,
\begin{equation}
    \text{SMAPE} = \frac{1}{P} \sum_{p=1}^P \left(\frac{100\%}{360} \sum_{t=1801}^{2160} \frac{|\bar{n}^p_t - n^p_t|}{|\bar{n}^p_t| + |n^p_t|} \right).
\end{equation} 
The predicted opinion volumes $\{\bar{n}^p_t\}$ are obtained using the \omm simulation algorithm. 
We 
%run the algorithm on \omm fitted on $\{n^p_{i,t} | t \in\ \mathcal{T}_{obs}\}$, then 
(1) condition on $\{n^p_{i,t} | t \in\ \mathcal{T}_{obs}\}$, 
(2) run the algorithm to sample $\{n^p_{i,t}\}$ on $\mathcal{T}_{pred}$, then 
(3) sum over opinion types $\{i\}$ to get predicted opinion volumes $n^p_{t} = \sum_i n^p_{i,t}$. 
We repeat $R=5$ times, and average over the samples to obtain $\{\bar{n}^p_t | t \in\ \mathcal{T}_{pred}\}$.

For opinion share prediction, we predict the opinion market shares $\{s^p_{i,t}\}$ for each platform $p$ on the evaluation period. 
To evaluate how well we predict opinion market shares, we calculate the KL divergence of predicted market shares $\{\bar{s}^p_t | t \in\ \mathcal{T}_{pred}\}$ (obtained similar to $\{\bar{n}^p_t\}$ described above) relative to actual market shares $\{s^p_t | t \in\ \mathcal{T}_{pred}\}$,
\begin{equation}
    \text{KL}^p(t) = \sum_{i=1}^M s^p_{i,t} \log \frac{\bar{s}^p_{i,t}}{s^p_{i,t}}.
\end{equation}

\textbf{Baselines.} 
We compare \omm with the discretized versions of the Correlated Cascades (CC) model \cite{Zarezade2017} and Competing Products (CP) model \cite{Valera2015} -- the current state-of-the-art models in product share modeling, covered in related works. 
For the bushfire study, we test the effectiveness of interventions by fitting \omm without $\{X_k(t)\}$ (indicated as $\omm \backslash \text{X}$). 

We also consider a feature-based predictive baseline -- the multivariate linear regression (MLR), used previously for online popularity prediction \citep{Pinto2013,Rizoiu2017}.
We build MLR with a one-week sliding window of three types of features: the previous event counts, exogenous signal $S(t)$ and interventions $\{X_k(t)\}$.
The predictive targets are the event counts $\{n_{i,t}^p\}$ for each point on $\mathcal{T}_{pred}$.
Analogous to \omm fitted without interventions $\{X_k(t)\}$, we additionally train MLR without $\{X_k(t)\}$ (indicated as MLR$\backslash \text{X}$) for the bushfire case study.

\omm, CC and CP are generative models typically designed for explainability and are known to be suboptimal for prediction \citep{Mishra2016}. 
In contrast, feature-driven approaches (e.g., MLR) use machine learning to predict using training features. 
Such approaches are designed mainly for prediction and have weaker explainability since they do not model the data-generation process \citep{Mishra2016}. 
In this work, we are interested in the dual tasks of predicting and explaining opinion market shares, hence our focus on generative approaches.

\begin{figure*}[!tp]
    \newcommand\myheight{0.32}
    \centering
    \subfigure[]{
        \includegraphics[height=\myheight\textheight]{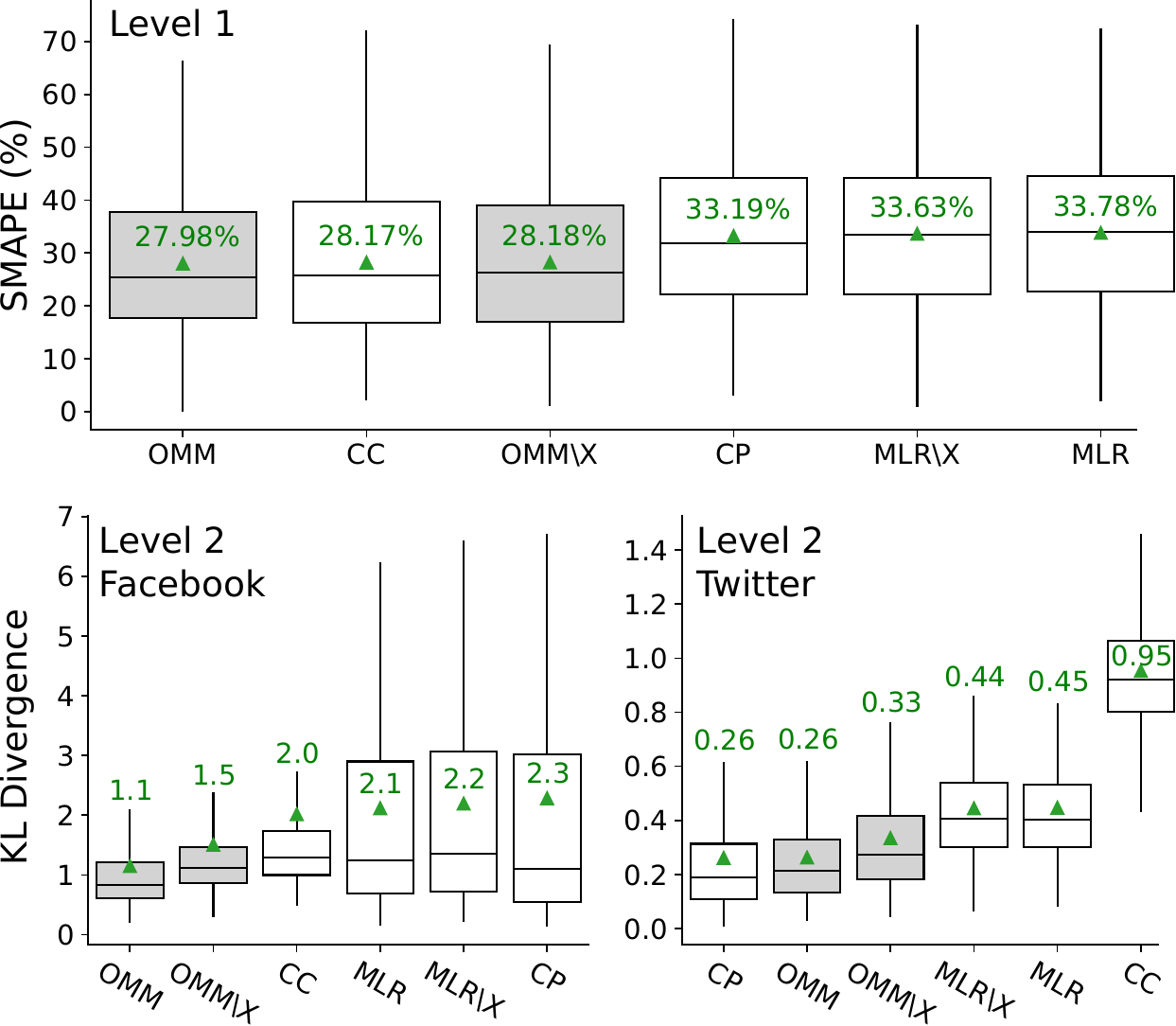}
        \label{fig:bushfire_summary}
    } 
    \subfigure[]{
        \includegraphics[height=\myheight\textheight]{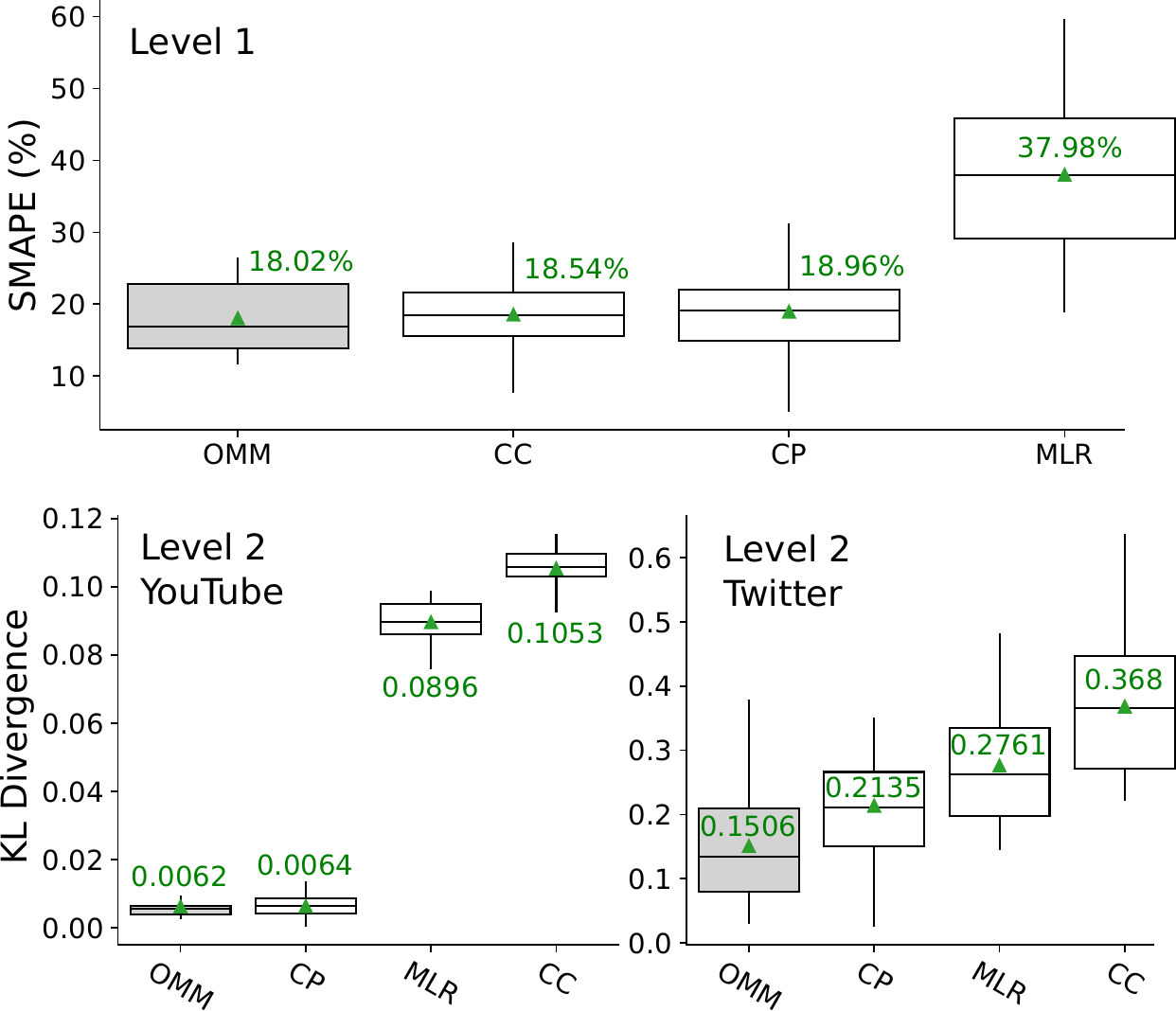}
        \label{fig:vevo_summary}
    } 
    \caption{
        Predictive evaluation of \omm on (a) Bushfire Opinions and (b) VEVO 2017 Top 10 datasets. 
        Boxplots are sorted left to right by the mean (shown with \textcolor{green}{green} triangle). 
        Shaded boxplots correspond to versions of \omm. 
        The top panels show the platform-averaged SMAPE of volumes on $\mathcal{T}_{pred}$.  
        Bottom panels plot the KL divergence of predicted and actual market shares.
    }
    \label{fig:performance_all}
\end{figure*}

\textbf{Predict Opinion Volumes.} 
\cref{fig:level1} showcases the observed (blue line) and modeled (orange line) opinion volumes for the bushfire dataset.
We visually observe that \omm achieves a tight fit on both the training and the prediction period (hashed area). 
The VEVO dataset results are shown in the online appendix \citep{appendix}.
% The modeled line shows the fit during 
We further compare \omm's predictive performances against baselines.
The top row of boxplots in \cref{fig:bushfire_summary,fig:vevo_summary} shows the platform-averaged SMAPE of predicted volumes for the bushfire and VEVO datasets, respectively. 
We make two observations.
First, in both case studies, \omm outperforms all baselines on opinion volume prediction.
Second, $\omm$ outperforms $\omm \backslash \text{X}$, indicating the role of media coverage in shaping attention.
% We see that , though CC slightly outperforms \omm without interventions for the bushfire dataset.

\textbf{Predict Opinion Market Share.} 
\cref{fig:level2} visualizes the observed (left column) and fitted during training and predicted during testing (right column) opinion market shares for the bushfire dataset.
We see that the opinion distribution on Twitter has significantly more variation than on Facebook, and that \omm closely captures the trend in opinion shares on both platforms.
The VEVO dataset results are in the online appendix \citep{appendix}. 
\cref{fig:bushfire_summary,fig:vevo_summary} show the KL-divergence of predicted market shares for the bushfire (Facebook and Twitter) and VEVO (YouTube and Twitter) datasets, respectively. 
% We additionally compare the performance of \omm with the CC and CP baselines (first and second boxplots) and test the effectiveness of interventions by implementing \omm with and without $\{X_k(t)\}$ (indicated as $\backslash \text{X}$).
We make several observations.
First, on the bushfire dataset, performance is better for Twitter than Facebook ($\text{KL}^{TW}(t) < \text{KL}^{FB}(t)$) due to Facebook having lower opinion counts than Twitter. 
Similarly, on the VEVO dataset $\text{KL}^{YT}(t) < \text{KL}^{TW}(t)$.
Second, \omm consistently outperforms all baselines on both datasets, except for Twitter on bushfires, where CP and \omm are comparable. 
CC performs poorly since it does not model asymmetric opinion interactions and assumes all opinions reinforce or inhibit one another. 
CP performs poorly on Facebook (Twitter) for the bushfire (VEVO) dataset due to CP not having the notion of limited total attention. 
Due to higher bushfire postings on Twitter, CP pays more attention to Twitter.
Lastly, \omm with $\{X_k(t)\}$ outperforms \omm without $\{X_k(t)\}$ on the bushfire dataset, suggesting that mainstream and controversial media effectively shape the opinion ecosystem.

\section{Interpreting \omm Elasticities}

In this section, we leverage the fitted \omm to uncover interactions across opinions and platforms in the bushfire dataset and artists in the VEVO dataset.

\begin{figure*}[!tp]
    \newcommand\myheight{0.28}
    \centering
    \subfigure[]{
        \includegraphics[height=\myheight\textheight]{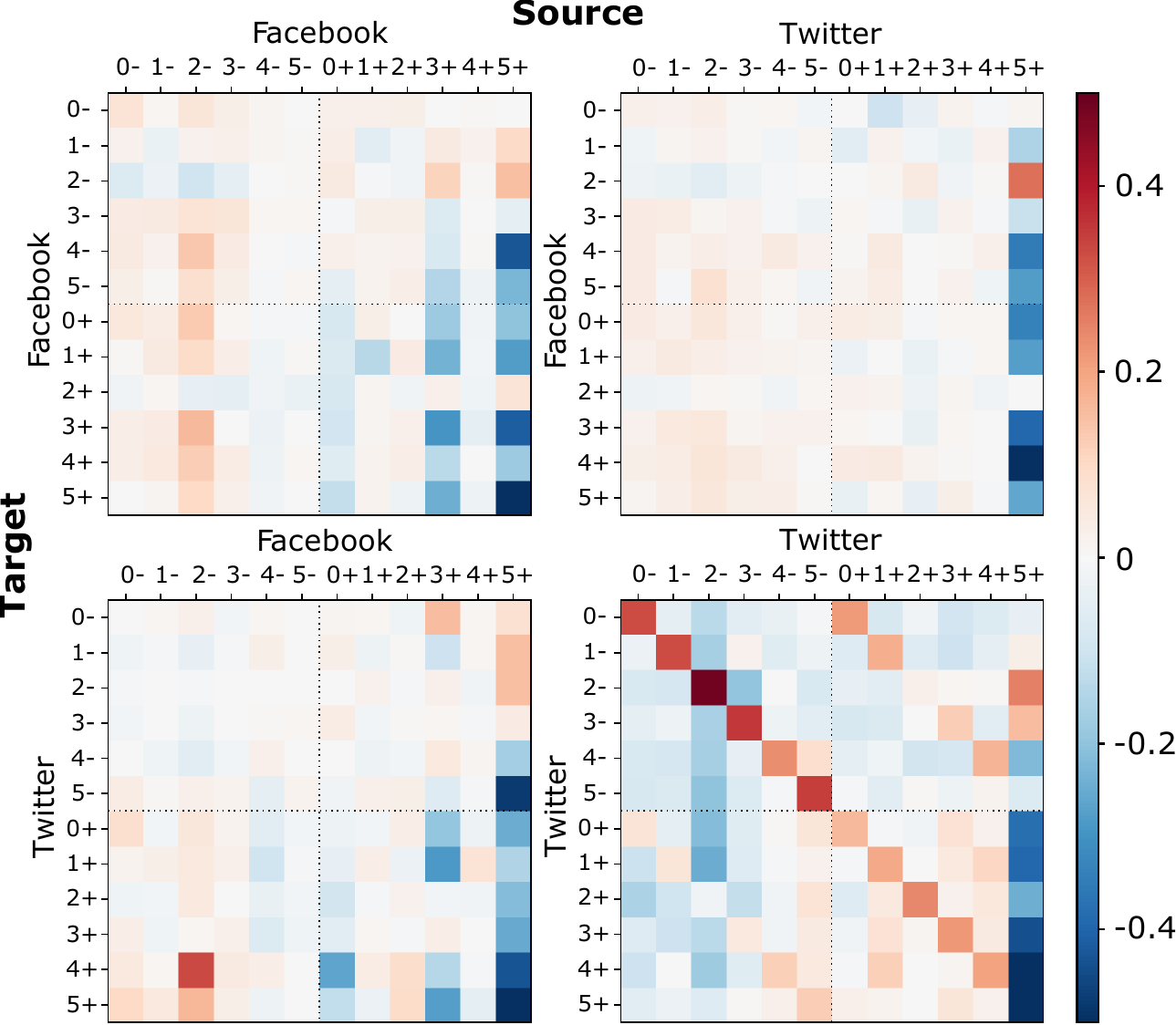}
        \label{fig:elasticity_endogenous}
    }%
    \subfigure[]{
        \includegraphics[height=\myheight\textheight]{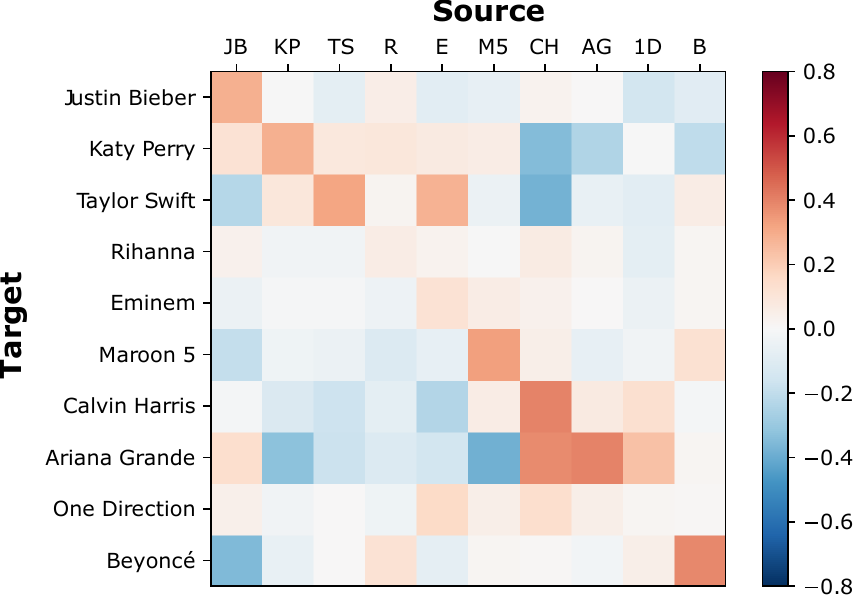}
        \label{fig:yt_level2}
    }%
    \caption{
        Interpretability of \omm. 
        (a) Endogenous elasticities $e(s^p_i(t), \lambda^q(t|j))$ across opinion pairs $(i,j)$ on respective platforms $(p,q)$ in the bushfire dataset. 
        Elasticities have direction and should be read from column (source) to row (target) for the platform and within each matrix.
        For example, the bottom-right matrix corresponds to influences from Twitter to Twitter; the cell $\{4-, 4+\}$ (\{row, column\}) is the influence of opinion $4+$ on $4-$, positive and large meaning that $4+$ has a strong reinforcing effect on $4-$.
        %(b) The intervention elasticities $e(s^p_i(t), \bar{X}(t))$ of the intervention on each opinion, on each platform.
        (b) YouTube elasticities $e(s^{YT}_i(t), \lambda^{YT}(t|j))$ across artist pairs $(i,j)$ in the VEVO 2017 Top 10 dataset.}
    \label{fig:interpretability}
\end{figure*}

\textbf{Uncovering Opinions Interactions.} 
To study opinion interactions in the bushfire dataset, we calculate the opinion share model elasticities (see \cref{eq:elasticity}) accounting for the endogenous volume $\lambda^p(t|j)$ and the intervention $\bar{X}_k(s)$ (see \cref{eq:opinionsharemodel_tendency}). 
The endogenous elasticities $e(s_i^p(t), \lambda^q(t|j))$ quantify the competition-cooperation interactions across opinions. 
The intervention elasticity $e(s_i^p(t), \bar{X}_k(t))$ quantifies the sensitivity of opinion market shares to intervention $X_k(t)$. 
We derive the elasticities and show results for $e(s_i^p(t), \bar{X}_k(t))$ in the online appendix \cite{appendix}. 
\cref{fig:elasticity_endogenous} reports the time averages of $e(s_i^p(t), \lambda^q(t|j))$.

First, we study intra-platform reinforcement (top-left \& bottom-right in \cref{fig:elasticity_endogenous}).
We see different behaviors for Facebook and Twitter.
%% MAR: I switched Twitter and Facebook to talk about the more exciting one first
For Twitter, we have two observations. 
First, there is strong self-reinforcement for opinions (i.e., main diagonal), indicative of the echo chamber effect \citep{Cinelli2021}. 
Second, there is significant cross-reinforcement among far-right sympathizers and opponents (i.e., diagonals on the upper-right \& lower-left submatrices), implying exchanges or arguments between opposing camps.
For Facebook, \omm detects little interaction among opinions, aside from the generally inhibitory effect of the opinions ``Australian bushfires and climate change are unrelated" (3+) and ``Bushfires are a normal summer occurrence" (5+) on other opinions.
This is because Facebook data was collected from far-right groups with limited interaction with users of the opposing side. 

\textbf{How to Effectively Suppress Far-Right Opinions.}
The above implies that confrontation is not the most effective method to suppress far-right opinions, as it has the potential to backfire by bringing even more attention to them.
A more effective method is boosting related counter-arguments; 
for instance, to suppress ``Australian bushfires were caused by random arsonists" (4+) on Twitter, \omm indicates to promote ``Climate change is real" (2-) and ``Greens are not the cause of the bushfires" (0-).
Boosting the opposite argument, i.e., ``Australian bushfires were not caused by random arsonists" (4-), would backfire. 
The opinion ``Bushfires are a normal summer occurrence in Australia" (5+) shows a different behavior: it reinforces most moderate opinions and inhibits far-right opinions. 
In particular, the ``Bushfires are normal" opinion (5+) appears to trigger ``Climate change is real" \mbox{(2-)}, probably due to the diametric opposition nature of these opinions. 
The effect of 5+ on 2- holds across every pair of platforms. Additionally, on Facebook, ``Australian bushfires and climate change are not related" (3+) has a similar effect on other opinions as the ``Bushfires are normal" opinion (5+), probably due to the similarity of their topic content.  

\textbf{Cross-Platform Reinforcement} is generally weak due to the Facebook far-right groups acting as a filter bubble.
Apart from the effect of ``Bushfires are normal" (5+) (see above), there is little cross-reinforcement among opinions from Twitter to Facebook. 
In the bottom-left matrix of \cref{fig:elasticity_endogenous}, we see that  ``Australian bushfires and climate change are not related" (3+) affects other opinions in a similar way to ``Bushfires are normal" (5+);
furthermore, ``Climate change is real" (2-) triggers ``Australian bushfires were caused by arsonists" (4+).

\textbf{Interactions Across VEVO Artists.} 
Lastly, in \cref{fig:yt_level2}, we shift our attention to the VEVO dataset and look at the YouTube-to-YouTube elasticities $e(s_i^{YT}(t), \lambda^{YT}(t|j))$ across our set of artists. 
The Twitter and cross-platform elasticities are available in the online appendix \cite{appendix}. 

We highlight three key observations. 
First, there is strong self-reinforcement for most artists (i.e., the main diagonal), an intuitive result reflecting these popular artists' strong fanbase. 
Second, \omm picks up non-trivial artist interactions that correspond with real-world events -- the animosity and friendship relations show up in their popularity dynamics. 
For instance, we see that Calvin Harris inhibits both Taylor Swift (the two broke up in 2016\footnote{\label{note2}\scriptsize people.com/celebrity/taylor-swift-calvin-harris-breakup-timeline/}) and Katy Perry (the two had a long-lasting feud\footnote{\label{note3}\scriptsize nme.com/news/music/katy-perry-ends-six-year-beef-calvin-harris-2128100}, due to Harris pulling out of Perry's 2011 tour last minute). 
Similarly, Taylor Swift and Justin Bieber have a mutually inhibiting relationship. 
The two have a well-known uneasy relationship\footnote{\label{note4}\scriptsize people.com/music/justin-bieber-selena-gomez-relationship-look-back/} since Justin Bieber and Selena Gomez used to date and the latter is one of Taylor Swift's close friends. 
Meanwhile, Calvin Harris and Ariana Grande have a reinforcing relationship, correlating with their collaboration ``Heatstroke'' released in March 2017. 
\omm picks up these relationships because we fit on online popularity driven by audience response. 
Fans of a given artist can choose to support or not support another artist based on real-world interactions, as indicated by the results above. 
Our third observation relates to the complexity of fanbase support for artists occupying the same genre: similar artists do not all just cooperate or compete for market share but can have unique pairwise relationships. 
For instance, Katy Perry, Taylor Swift and Ariana Grande occupy a similar niche (mainstream pop). 
However, our model uncovers that Taylor Swift and Katy Perry reinforce each other, while these two inhibit (and are inhibited by) Ariana Grande.
% \begin{figure*}[!tbp]
%     \newcommand\myheight{0.205}
%     \centering
%     \subfigure[]{
%         \includegraphics[height=\myheight\textheight]{plots/fig5/youtube_level1.pdf}
%         \label{fig:yt_level1}
%     } 
%     \subfigure[]{
%         \includegraphics[height=\myheight\textheight]{plots/fig5/youtube_level2.pdf}
%         \label{fig:kl}
%     } 
%     \subfigure[]{
%         \includegraphics[height=\myheight\textheight]{plots/fig5/youtube_sourcetarget_v2.pdf}
%         \label{fig:yt_level2}
%     } 
%     \caption{Evaluating \omm on the VEVO dataset. We train \omm on days 1-75 and predict on days 76-100 (shaded area). We present results for YouTube and Twitter, respectively. 
%     (a) Actual (dashed \textcolor{blue}{blue} lines) vs. fitted/predicted  (\textcolor{orange}{orange} lines) volumes; 
%     (b) KL-divergence on the evaluation set for CC, CP baselines and \omm; and 
%     (c) time-averaged endogenous YouTube elasticities $e(s^{YT}_i(t), \lambda^{YT}(t|j))$ across artist pairs $(i,j)$. 
%     Elasticities are directional and are read from column (source) to row (target).}
%     \label{fig:youtube}
% \end{figure*}

\section{\omm as a Testbed for Interventions}
\label{subsec:what-if}

The interventions $\{X_k(t)\}$ can lead to delayed effects in the opinion ecosystem due to the opinion dependency structure.
For example, if an intervention is designed to boost a target opinion, it will indirectly boost all other opinions with a cooperative relationship with the target opinion. 
Furthermore, it will inhibit opinions with a competitive relationship with the target.
Since elasticities only inform us of the \textit{instantaneous} effect on opinion market shares, we perform a what-if exercise to study the role of interventions in the bushfire case study.
We vary the size of the intervention and synthetically sample outcomes to observe the long-term effects of media coverage on the opinion ecosystem. 

\textbf{What-if can inform A/B test design.}
We train \omm on observational data; therefore, the inferred effects of interventions $\{X_k(t)\}$ are not causal impact estimates but rather evidence of causal effects. 
However, the previous section demonstrates that \omm can uncover complex relationships across opinions, providing compelling evidence that \omm is also able to uncover relationships between opinions and interventions.
Therefore, the what-if exercise in this section showcases \omm as a testbed for interventions, usable for designing A/B testing that determines true causal effects. 
The \omm informs us of the effectiveness of interventions, allowing us to prioritize which specific interventions to test.

\textbf{``What-if'' Setup.}
We test the effect of interventions by synthetically increasing or decreasing their volumes past a given time point (see top panel of \cref{fig:teaser}) and measuring the percentage change in far-right opinions.
% We observe the change in far-right opinion shares when we modulate interventions one at a time, keeping the others constant.
% 
Let $k^* \in \{1, \ldots, K\}$ be the index of the modulated intervention. 
We modulate $X_{k^*}(t)$ as $X_{k^*}^{(r)}(t) = X_{k^*}(t) + r \cdot \mu_{X_{k^*}} \cdot \mathds{1}_{(t > 1800)}$, where $\mathds{1}_{(\cdot)}$ is the indicator function and $\mu_{X_{k^*}}$ is the mean volume of $X_{k^*}
(t)$ on $\mathcal{T}_{obs}$. 
The parameter $r$ controls the percent increase ($r>0$) or decrease ($r<0$) in media coverage beyond the change point $t=1800$; 
$r=0$ is the original $X_{k^*}(t)$.
We run \omm with $X^{(r)}_{k^*}(t)$ for various $r$, and keep $X_{k}(t)$ fixed for $k \neq k^*$.
We quantify the effects of intervention $X_{k^*}(t)$ as
the average percent change (relative to $r=0$) in the opinion market shares after the change point, i.e., $\mathcal{T}_{pred}$. 
We perform this procedure for all $k^* \in \{1, \ldots, K\}$.

\begin{figure}[!tp]
    \centering
    \includegraphics[width=0.47\textwidth]{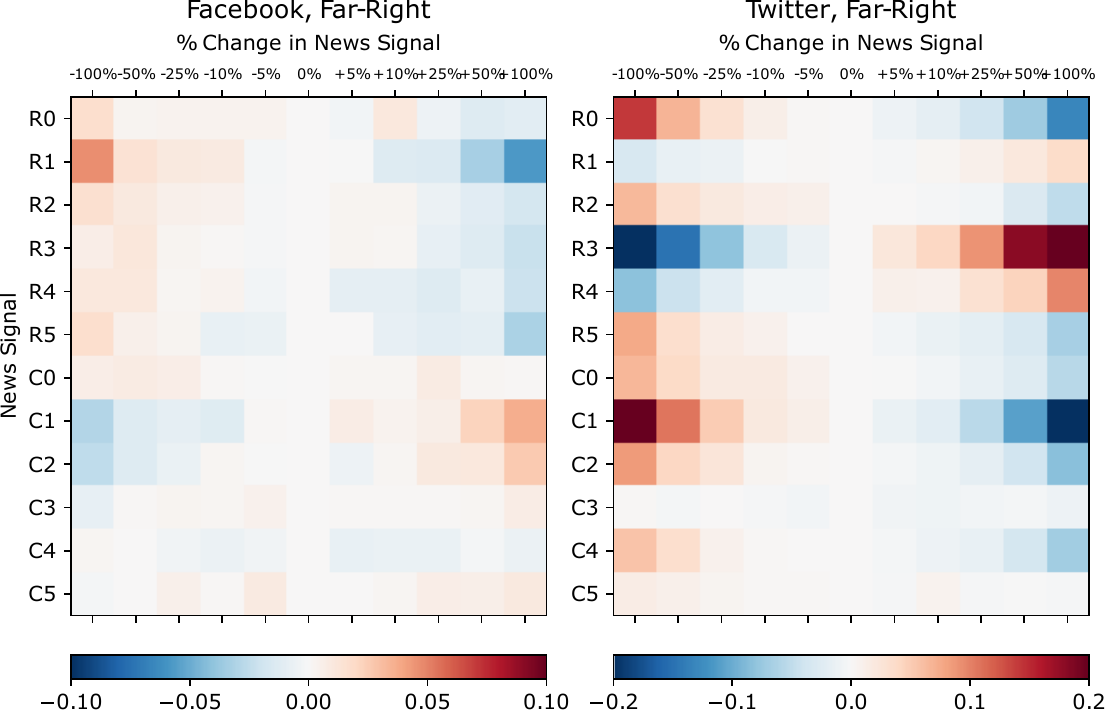}
    \caption{
        We modulate the volume of reputable (R) and controversial (C) news for each opinion (in $\{0,1,2,3,4,5\}$) from $-100\%$ to $100\%$ of the mean volume and simulate \omm to see the percent change in the far-right (+) opinion market shares on Facebook (left) and Twitter (right).}
	\label{fig:whatif}
\end{figure}

\textbf{How News Influences Far-Right Opinions.}
\cref{fig:whatif} shows the average percent changes in the market share of far-right opinions when modulating the interventions $\{R_i(t), C_i(t)\}$ one at a time for various $r$ over $50$ simulations. 
On Facebook, far-right opinions are suppressed by reputable news and reinforced by the majority of controversial news, except for news concerning ``Greens policies are the cause of the Australian bushfires" ($R_0$) and ``Australian bushfires were caused by arsonists" ($R_4$).
On Twitter, both reputable and controversial news suppress far-right opinions, except for reputable news concerning ``Australian bushfires/climate change are unrelated" ($R_3$), ``Australian bushfires were caused by arsonists" ($R_4$) and to a lesser extent ``Mainstream media cannot be trusted" ($R_1$). 

We have two key insights. 
First, we see that the effect of the news on Facebook is modest compared to Twitter since the far-right public groups on Facebook behave as almost perfect filter bubbles in which news has little penetration. 
Second, indiscriminately increasing reputable news is not an effective strategy for suppressing far-right opinions on Twitter (see $R_3$ and $R_4$). 
Doing so can backfire since it brings even more attention to far-right users and their narratives \cite{Peucker2022}.
% often link and co-opt such news articles to suit their ideology

\textbf{How to effectively use the testbed.}
Assuming that A/B testing is performed by an entity in control of reputable news coverage ($R_i$ here above), the results above indicate that the test should mainly concentrate on the effects of increasing $R_1$ (on Facebook), increasing $R_0$ and decreasing $R_3$ and $R_4$ (on Twitter).
We leave as future work the design and execution of such an experiment.
Our analysis in this paper focuses on mitigating far-right opinions with media coverage.
However, \omm can be leveraged as an intervention evaluation tool for information operations in other domains and fighting mis- \& disinformation and online propaganda.

% \section{The attention dynamics of music artists}
% \label{subsection:vevo}

% In this section, we showcase the generalizability of \omm by fitting on another dataset: the YouTube and Twitter attention volumes for the most popular artists on the VEVO platform in 2017.
% We demonstrate that \omm can uncover the latent relationships among artists by modeling their joint popularity dynamics. 
% First, we describe the \textit{VEVO 2017 Top 10} dataset and we define the set of exogenous signals $\{S_i(t)\}$.
% Next, we evaluate the predictive performance of \omm and demonstrate interpretability of the model elasticities.
% %  (\cref{subsection:vevo_prediction}).

\section{Summary and Discussion}
This work introduces the Opinion Market Model (\omm), a novel two-tier model of the dynamics of the online opinion ecosystem. 
The first tier models the size of the attention market, and the second tier models opinions competing or cooperating for limited public attention under the influence of positive interventions. 
We develop algorithms to simulate and estimate \omm, showing the convergence using synthetic data. 
We demonstrate real-world applicability on a dataset of Facebook and Twitter discussions containing moderate and far-right opinions on bushfires and climate change \cite{Kong2021} and a dataset of YouTube and Twitter attention volumes for popular artists on VEVO \cite{Wu2019}. 
We show \omm predicts opinion market shares better than state-of-the-art baselines \cite{Valera2015,Zarezade2017} and uncovers latent competitive and cooperative interactions across opinions: self-reinforcement attributable to the echo chamber effect and interactions between far-right sympathizers and opponents.
Lastly, we quantify the effect of reputable and controversial media coverage on Facebook and Twitter.

\textbf{Scope of Study.} 
This work focuses on the manifestation of far-right opinions in the context of the 2019-2020 Australian bushfires. 
Note that far-right ideology manifests in other political issues (e.g., gun control, LGBT rights, xenophobia), which we do not tackle here. 
Moreover, we do not focus on the general political science of far-right ideology since we are projecting onto a specific context.

% \subsection*{Acknowledgments}
\paragraph{Acknowledgments.}
This work was partially funded by the Australian Department of Home Affairs, the Defence Science and Technology Group, the Defence Innovation Network and the Australian Academy of Science.

\bibliography{aaai24}



\section*{Supplementary material}

# Appendix:
# Opinion Market Model: Stemming Far-Right Opinion Spread Using Positive Interventions

Pio Calderon, Rohit Ram, Marian-Andrei Rizoiu

This document accompanies the submission *Opinion Market Model: Stemming Far-Right Opinion Spread Using Positive Interventions*. The information in this document complements the submission and is presented here for completeness reasons. It is not required to understand the main paper or reproduce the results.

## A  Full Table of Notations

Table 1 shows the full table of notations for the OMM.

## B  Model Likelihood, Estimation, Simulation and Gradients

In this section we provide technical details for main text Section 3. We first go over the derivation of the model likelihoods, followed by the estimation and simulation algorithms for the two-tier OMM model. Lastly, we derive the model gradients.

**Likelihood formulation**

**Likelihood function** $\mathcal{L}_1(\Theta_1|\{n_t^p\}_{p,t})$, where $\Theta_1 = \{\mu^P, \alpha^{pq}, \theta\}$. The log-likelihood function can be derived by

$$\mathcal{L}_1(\Theta_1|\{n_t^p\}_{p,t})$$
$$= \log \mathbb{P}\left\{\bigcup_{t=1}^{T}\bigcup_{p=1}^{P}[N^p(t) = n_t^p]\right\}$$
$$= \sum_{t=1}^{T}\sum_{p=1}^{P} \log \mathbb{P}\left\{N^p(t) = n_t^p\right\}$$
$$= \sum_{t=1}^{T}\sum_{p=1}^{P} \log\left[\frac{e^{-\lambda^p(t)}\lambda^p(t)^{n_t^p}}{n_t^p!}\right]$$
$$\propto \sum_{t=1}^{T}\sum_{p=1}^{P}\left[n_t^p \log \lambda^p(t) - \lambda^p(t)\right] \quad (1)$$

**Likelihood function** $\mathcal{L}_2(\Theta_2|\Theta_1, \{n_{i,t}^p\}_{i,p,t})$, where $\Theta_2 = \{\mu_j^p, \gamma_{ik}^p, \beta_{ij}^{pq}\}$. Instead of estimating the parameters $\mu_j^p \in \mathbb{R}$, we can estimate the normalized parametes $\hat{\mu}_j^p \in [0, 1]$, where $\mu_j^p = \mu^p \cdot \hat{\mu}_j^p$. Given that the magnitudes of $\gamma_{ik}^p$ and $\beta_{ij}^{pq}$ are typically less than one, estimating normalized parameters $\hat{\mu}_j^p$ instead of $\mu_j^p$ avoids scaling problems. Hence, we optimize for $\Theta_2 = \{\hat{\mu}_j^p, \gamma_{ik}^p, \beta_{ij}^{pq}\}$.

$$\mathcal{L}_2(\Theta_2|\Theta_1, \{n_{i,t}^p\}_{i,p,t})$$
$$= \log \mathbb{P}\left\{\bigcup_{t=1}^{T}\bigcup_{p=1}^{P}\bigcup_{i=1}^{M}[N_i^p(t) = n_{i,t}^p]\right\}$$
$$= \sum_{t=1}^{T}\sum_{p=1}^{P}\sum_{i=1}^{M} \log \mathbb{P}\left\{N_i^p(t) = n_{i,t}^p\right\}$$
$$= \sum_{t=1}^{T}\sum_{p=1}^{P}\sum_{i=1}^{M} \log \left[\frac{e^{-\lambda_i^p(t)}\lambda_i^p(t)^{n_{i,t}^p}}{n_{i,t}^p!}\right]$$
$$\propto \sum_{t=1}^{T}\sum_{p=1}^{P}\sum_{i=1}^{M}\left[n_{i,t}^p \log \lambda_i^p(t) - \lambda_i^p(t)\right]$$
$$= \sum_{t=1}^{T}\sum_{p=1}^{P}\sum_{i=1}^{M}\left[n_{i,t}^p \log(\lambda^p(t) \cdot s_i^p(t)) - (\lambda^p(t) \cdot s_i^p(t))\right]$$
$$= \sum_{t=1}^{T}\sum_{p=1}^{P}\sum_{i=1}^{M}\left[n_{i,t}^p (\log \lambda^p(t) + \log s_i^p(t)) - (\lambda^p(t) \cdot s_i^p(t))\right]$$
(2)

**Estimation algorithm.**

We estimate the parameters of OMM with the following two-step formula:

1. Given $\{n_t^p\}_{p,t}$, find $\hat{\Theta}_1 = \Theta_1$ that maximizes
$$\mathcal{L}_1(\Theta_1|\{n_t^p\}_{p,t}) = \sum_{p,t}[n_t^p \log \lambda^p(t) - \lambda^p(t)]. \quad (3)$$

2. Given $\{n_{i,t}^p\}_{i,p,t}$ and $\hat{\Theta}_1$, find $\hat{\Theta}_2 = \Theta_2$ that maximizes
$$\mathcal{L}_2(\Theta_2|\hat{\Theta}_1, \{n_{i,t}^p\}_{i,p,t}) =$$
$$\sum_{i,p,t}[n_{i,t}^p(\log \lambda^p(t) \cdot s_i^p(t)) - (\lambda^p(t) \cdot s_i^p(t))]. \quad (4)$$

Due to the non-convexity of $\mathcal{L}_1(\cdot)$ and $\mathcal{L}_2(\cdot)$, we avoid local maxima by running the algorithm for multiple starting

| Notation | Interpretation |
|---|---|
| $P$ | number of social media platforms |
| $M$ | number of opinion types |
| $K$ | number of positive interventions |
| $T$ | terminal time |
| **Variable** | |
| $S(t)$ | input signal accounting for the volume of exogenous events |
| $X_k(t)$ | input signal corresponding to the $k^{th}$ positive intervention |
| $s_i^p(t)$ | market share of opinion $i$ on platform $p$ at time $t$ |
| $\lambda^p(t)$ | conditional intensity of attention volume (Opinion Volume Model) |
| $\lambda^p(t|i)$ | conditional intensity of opinion $i$, assuming independence of opinions |
| $\lambda_i^p(t)$ | conditional intensity of opinion $i$ (Opinion Share Model) |
| $N^p(t)$ | total attention volume on platform $p$ at time $t$, based on OMM |
| $N_i^p(t)$ | number of posts with opinion $i$ on platform $p$ at time $t$, based on OMM |
| $e(s_i^p(t), \lambda^q(t|j))$ | opinion share model elasticity w.r.t. endogenous dynamics |
| $e(s_i^p(t), X_k(t))$ | opinion share model elasticity w.r.t. intervention |
| **Data** | |
| $n_t^p$ | number of posts on platform $p$ at time $t$ |
| $n_{i,t}^p$ | number of posts on platform $p$ with opinion $i$ at time $t$ |
| $s_{i,t}^p$ | fraction of posts on platform $p$ with opinion $i$ at time $t$ |
| **Parameter** | |
| $\mu_j^p$ | exogenous scaling term for opinion $j$ on platform $p$, $\mu_j^p$ |
| $\mu^p$ | exogenous scaling term for platform $p$, given by $\mu^p = \sum_{j=1}^M \mu_j^p$ |
| $\alpha^{pq}$ | excitation parameter for intra-platform ($p=q$) and inter-platform (for $p \neq q$) dynamics |
| $\theta$ | memory parameter, describing how fast an event is forgotten, $\theta \in [0,1]$ |
| $\gamma_{ik}^p$ | measure of the direct effect of the $k^{th}$ intervention on the market share of opinion $i$ on platform $p$ |
| $\beta_{ij}^{pq}$ | measure of the direct effect that opinion $j$ on platform $q$ has on the market share of opinion $i$ on platform $p$. |

Table 1: Table of notations.

points and selecting the combination with the largest likelihood.

### Sampling algorithm.

We generate samples from OMM by looping the following steps over $t \in \{1, \ldots, T\}$ and each platform $p$ and opinion $i$.

1. Compute $\lambda_i^p(t) = \lambda^p(t| \cup_{q,s<t} \{n_s^q\}) \cdot s_i^p(t| \cup_{q,j,s<t} \{n_{j,s}^q\})$.

2. Draw a sample $n_{i,t}^p \sim \text{Poi}(\lambda_i^p(t))$.

### Gradients

**Gradient** $\partial_{\Theta_1} \mathcal{L}_1(\Theta_1|\{n_t^p\}_{p,t})$. Differentiating Eq. (3), we get

$$\partial_{\Theta_1} \mathcal{L}_1(\Theta_1|\{n_t^p\}_{p,t}) = \sum_{t=1}^T \sum_{p=1}^P \frac{\partial_{\Theta_1} \lambda^p(t)}{\lambda^p(t)} \cdot [n_t^p - \lambda^p(t)], \quad (5)$$

where

$$\partial_{\mu^q} \lambda^p(t) = \delta_{pq} \cdot S(t) \quad (6)$$

$$\partial_{\alpha^{qr}} \lambda^p(t) = \delta_{pq} \cdot \sum_{s<t} f(t-s) \cdot N^r(s) \quad (7)$$

$$\partial_\theta \lambda^p(t) = \sum_{q=1}^P \sum_{s<t} \alpha^{pq} \cdot \partial_\theta f(t-s) \cdot N^q(s) \quad (8)$$

$$\partial_\theta f(t) = (1-\theta)^{t-2} [1 - \theta t]. \quad (9)$$

**Gradient** $\partial_{\Theta_2} \mathcal{L}_2(\Theta_2|\Theta_1, \{n_{i,t}^p\}_{i,p,t})$. Differentiating Eq. (4), we get

$$\partial_{\Theta_2} \mathcal{L}_2(\Theta_2|\Theta_1, \{n_{i,t}^p\}_{i,p,t}) =$$
$$\sum_{t=1}^T \sum_{p=1}^P \sum_{i=1}^M \left[ \frac{\partial_{\Theta_2} \lambda^p(t)}{\lambda^p(t)} + \frac{\partial_{\Theta_2} s_i^p(t)}{s_i^p(t)} \right]$$
$$\cdot \left[ n_{i,t}^p - \lambda^p(t) \cdot s_i^p(t) \right], \quad (10)$$

where upon differentiating main text Eq. (7) and main text Eq. (8) we have

$$\partial_{\Theta_2} s_i^p(t) = \frac{\left[\sum_j \mathcal{A}_j^p(t)\right] \partial_{\Theta_2} \mathcal{A}_i^p(t) - \mathcal{A}_i^p(t) \left[\sum_j \partial_{\Theta_2} \mathcal{A}_j^p(t)\right]}{\left[\sum_j \mathcal{A}_j^p(t)\right]^2}, \quad (11)$$

and

$$\partial_{\Theta_2} \mathcal{A}_i^p(t) = \mathcal{A}_i^p(t) \cdot \partial_{\Theta_2} \mathcal{T}_i^p(t). \quad (12)$$

Plugging in Eq. (12) into Eq. (11), we get

$$\partial_{\mathbf{\Theta}_2} s_i^p(t) = \frac{\left[\sum_j \mathcal{A}_j^p(t)\right] \mathcal{A}_i^p(t) \cdot \partial_{\mathbf{\Theta}_2} \mathcal{T}_i^p(t)}{\left[\sum_j \mathcal{A}_j^p(t)\right]^2}$$
$$- \frac{\mathcal{A}_i^p(t) \left[\sum_j \mathcal{A}_i^p(t) \cdot \partial_{\mathbf{\Theta}_2} \mathcal{T}_j^p(t)\right]}{\left[\sum_j \mathcal{A}_j^p(t)\right]^2}$$
$$= \frac{\mathcal{A}_i^p(t) \left[\sum_j \mathcal{A}_j^p(t) \cdot \partial_{\mathbf{\Theta}_2} \mathcal{T}_i^p(t)\right]}{\left[\sum_j \mathcal{A}_j^p(t)\right]^2}$$
$$- \frac{\left[\sum_j \mathcal{A}_i^p(t) \cdot \partial_{\mathbf{\Theta}_2} \mathcal{T}_j^p(t)\right]}{\left[\sum_j \mathcal{A}_j^p(t)\right]^2}$$
$$= \frac{\mathcal{A}_i^p(t)}{\left[\sum_j \mathcal{A}_j^p(t)\right]^2} \cdot \sum_j \mathcal{A}_j^p(t) \cdot \left[\partial_{\mathbf{\Theta}_2} \mathcal{T}_i^p(t) - \partial_{\mathbf{\Theta}_2} \mathcal{T}_j^p(t)\right]$$
$$= s_i^p(t) \cdot \sum_j s_j^p(t) \cdot \left[\partial_{\mathbf{\Theta}_2} \mathcal{T}_i^p(t) - \partial_{\mathbf{\Theta}_2} \mathcal{T}_j^p(t)\right],$$

and so

$$\frac{\partial_{\mathbf{\Theta}_2} s_i^p(t)}{s_i^p(t)} = \sum_j s_j^p(t) \cdot \left[\partial_{\mathbf{\Theta}_2} \mathcal{T}_i^p(t) - \partial_{\mathbf{\Theta}_2} \mathcal{T}_j^p(t)\right]$$
$$= \partial_{\mathbf{\Theta}_2} \mathcal{T}_i^p(t) - \sum_j s_j^p(t) \cdot \partial_{\mathbf{\Theta}_2} \mathcal{T}_j^p(t)$$
$$= \sum_j (\delta_{ij} - s_j^p(t)) \cdot \partial_{\mathbf{\Theta}_2} \mathcal{T}_j^p(t) \quad (13)$$

Plugging in Eq. (13) into Eq. (10), we have

$$\partial_{\mathbf{\Theta}_2} \mathcal{L}_2(\mathbf{\Theta}_2|\mathbf{\Theta}_1, \{n_{i,t}^p\}_{i,p,t}) = \sum_{t=1}^T \sum_{p=1}^P \sum_{i=1}^M \left[\frac{\partial_{\mathbf{\Theta}_2} \lambda^p(t)}{\lambda^p(t)} \right.$$
$$\left. + \sum_{j=1}^M (\delta_{ij} - s_j^p(t)) \cdot \partial_{\mathbf{\Theta}_2} \mathcal{T}_j^p(t)\right] \cdot \left[n_{i,t}^p - \lambda^p(t) \cdot s_i^p(t)\right],$$
(14)

where

$$\partial_{\mu_j^q} \lambda^p(t) = \delta_{pq} \cdot S(t) \quad (15)$$
$$\partial_{\gamma_{ik}^q} \lambda^p(t) = 0 \quad (16)$$
$$\partial_{\beta_{ij}^{pq}} \lambda^p(t) = 0 \quad (17)$$
$$\partial_{\mu_j^q} \mathcal{T}_i^p(t) = \beta_{ij}^{pq} \quad (18)$$
$$\partial_{\gamma_{jk}^q} \mathcal{T}_i^p(t) = \delta_{ij} \delta_{qp} \cdot \sum_{s<t} f(t-s) \cdot X_k(s) \quad (19)$$
$$\partial_{\beta_{jk}^{qr}} \mathcal{T}_i^p(t) = \delta_{ij} \delta_{qp} \cdot \lambda^r(t|k) \quad (20)$$

**Fitting on multiple samples.**
Suppose that we are given $n_{samples}$ samples to fit the OMM.

Let $\mathcal{S} = \{\{n_{i,t}^p\}_{i,p,t}^s | s \in \{1, \ldots, n_{samples}\}\}$. One can define the joint likelihood over $\mathcal{S}$ as the average likelihood over the $n_{samples}$ samples. That is,

$$\mathcal{L}_1(\mathbf{\Theta}_1|\mathcal{S}) = \frac{1}{n_{samples}} \sum_{s=1}^{n_{samples}} \mathcal{L}_1(\mathbf{\Theta}_1|\{n_t^p\}_{p,t}^s)$$

$$\mathcal{L}_2(\mathbf{\Theta}_2|\mathbf{\Theta}_1, \mathcal{S}) = \frac{1}{n_{samples}} \sum_{s=1}^{n_{samples}} \mathcal{L}_2(\mathbf{\Theta}_2|\mathbf{\Theta}_1, \{n_{i,t}^p\}_{i,p,t}^s).$$

where $\mathcal{L}_1(\cdot)$ and $\mathcal{L}_2(\cdot)$ are defined in Eq. (3) and Eq. (4), respectively. Parameter optimization proceeds in the same setup as the two-step procedure detailed in Section 3.

## C  Additional Results for Synthetic Data

In Fig. 1 we show the behavior of the RMSE for $\boldsymbol{\mu}$, $\boldsymbol{\theta}$, $\boldsymbol{\beta}$ as we increase the training time $T$. Error stabilises and the model converges as we increase $T$. In Fig. 2 we show behavior of the RMSE of our parameters as we vary the number of samples in the joint fit $n_{samples}$. Increasing the number of samples improves performance on the first-tier parameters $\boldsymbol{\mu}$, $\boldsymbol{\theta}$ and $\boldsymbol{\alpha}$, but does not have a strong improvement on the second-tier parameters $\boldsymbol{\beta}$ and $\boldsymbol{\gamma}$. Increasing $n_{samples}$ stabilizes the likelihood of the fit.

## D  Additional Model Details
**Stability of the softmax function**

In main text Eq. (8), the tendency $\mathcal{T}_i^p(t)$ is unconstrained, and it can take both really large or really small numbers, which leads to numerical overflow and underflow in main text Eq. (7). To remedy this, instead of main text Eq. (8) we use

$$\tilde{\mathcal{A}}_i^p(t) = \exp\left[\mathcal{T}_i^p(t) - \max_{k \in \{1,\ldots,M\}} \mathcal{T}_k^p(t)\right],$$

which does not affect market share calculations since

$$s_i^p(t) = \frac{\mathcal{A}_i^p(t)}{\sum_{j=1}^M \mathcal{A}_j^p(t)} = \frac{\tilde{\mathcal{A}}_i^p(t)}{\sum_{j=1}^M \tilde{\mathcal{A}}_j^p(t)}.$$

Gradient and elasticity calculations are unaffected when we use $\tilde{\mathcal{A}}_i^p(t)$ instead of $\mathcal{A}_i^p(t)$.

**Bushfire opinion share model regularization**

Fitting the opinion share model to data involves estimation of $\mathbf{\Theta}_2 = \{\hat{\mu}_j^p, \gamma_{ik}^p, \beta_{ij}^{pq}\}$, a total of $P \times M + P \times M \times K + P^2 \times M^2$ parameters. Given the high dimensionality of this space, for the bushfire case study we opted to reduce the space of solutions by imposing platform-dependent structure on $\gamma_{ik}^p$ via regularization.

Let $\mathbf{M}^p$ be the mask matrices given by

$$M_{ik}^{FB} = \begin{cases} 0, & \left(i \leq \lfloor \frac{K}{2} \rfloor \wedge k \leq \lfloor \frac{K}{2} \rfloor\right) \vee \left(i > \lfloor \frac{K}{2} \rfloor \wedge k > \lfloor \frac{K}{2} \rfloor\right) \\ 1, & \text{otherwise} \end{cases}$$

$$M_{ik}^{TW} = \begin{cases} 0, & (i = k) \vee \left(i = k - \lfloor \frac{K}{2} \rfloor\right) \vee \left(k = i - \lfloor \frac{K}{2} \rfloor\right) \\ 1, & \text{otherwise} \end{cases}$$

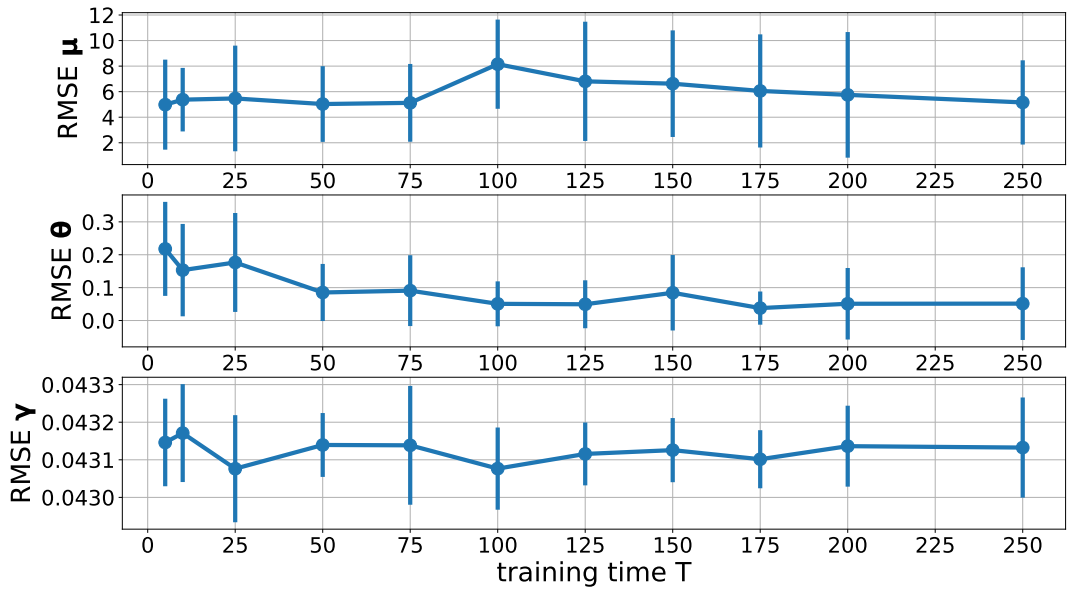

Figure 1: Additional results on synthetic data. We show the convergence of the RMSE of the $\boldsymbol{\mu}, \theta, \boldsymbol{\beta}$ as we increase the training time $T$.

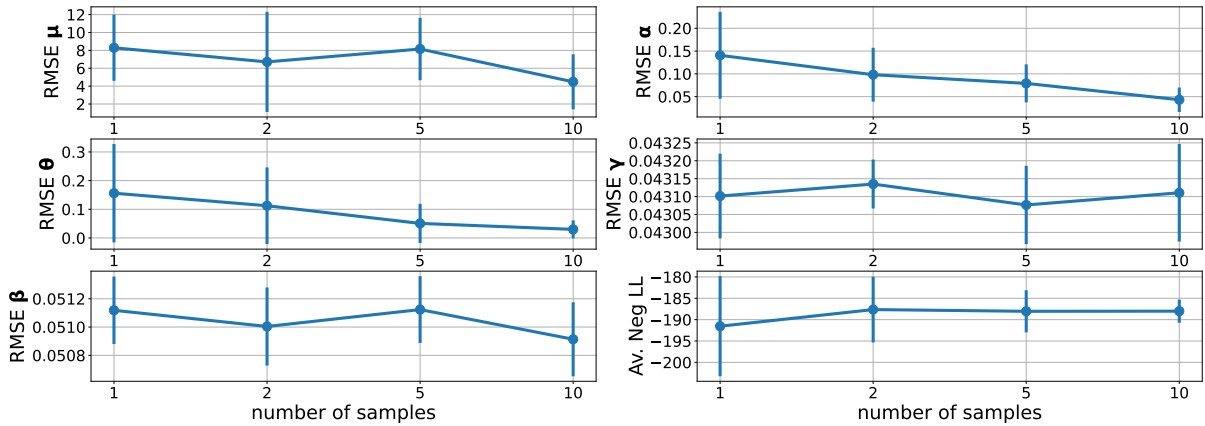

Figure 2: Additional results on synthetic data. We show the behavior of the RMSE of our parameter set and the average negative log likelihood as we vary the number of samples in the joint fit.

Instead of Eq. (4), we solve

$$\hat{\Theta}_2 = \arg\min_{\Theta_2} \left[ -\mathcal{L}_2(\Theta_2|\hat{\Theta}_1, \{n^p_{i,t}\}_{i,p,t}) + \lambda \sum_{p,i,k} |\gamma^p_{ik} \cdot M^p_{ik}| \right],$$

where $\lambda$ is a regularization parameter we set to 0.1.

Intuitively, the regularization encodes the echo chamber effect observed in Facebook far-right groups: far-right sympathizers interact mostly with news from controversial outlets, with limited interaction with reputable outlets. Similarly, far-right opponents interact mostly with reputable news, with limited interaction with controversial outlets. Given the more dialog-heavy nature of Twitter where exchanges between sympathizers and opponents are more common, we assume news from reputable and controversial outlets penetrate both far-right sympathizers and opponents, though we assume that sympathizers and opponents of a given opinion are only concerned with (and influenced by) news of the same opinion.

**Transformations on $\lambda^q(t|j)$ and $\bar{X}_k(s)$ in $\mathcal{T}_i^p(t)$**

We perform two transformations on $\lambda^q(t|j)$ and $\bar{X}_k(s)$ in main text Eq. (9) to improve model fit.

First, it was observed that $\lambda^q(t|j)$ has a skewed distribution over time. The skewness is problematic since we estimate a time-independent linear parameter $\beta^{pq}_{ij}$ for the direct effect of $\lambda^q(\cdot|j)$ on $\mathcal{T}_i^p(t)$. To reduce the skewness of $\lambda^q(\cdot|j)$, we transform $\lambda^q(t|j)$ to $\log[\lambda^q(t|j)+1]$, where we add 1 to avoid taking the logarithm of 0. Second, since $\mathcal{T}_i^p(t)$ is a linear combination of $\lambda^q(t|j)$ and $\bar{X}_k(t)$ terms, which could have totally different scales, we standardize these terms to bring them to a normalized scale. Let

$$\tilde{\lambda}^q(t|j) = \frac{\log[\lambda^q(t|j)+1] - \mathrm{mean}_s \log[n^q_{j,s}+1]}{\mathrm{std}_s \log[n^q_{j,s}+1]},$$

$$\tilde{X}_k(t) = \frac{\bar{X}_k(t) - \mathrm{mean}_s[\bar{X}_k(s)]}{\mathrm{std}_s[\bar{X}_k(s)]}$$

where the mean and standard deviation are computed over the training period.

Instead of main text Eq. (9), we use the following form of the tendency:

$$\mathcal{T}_i^p(t) = \sum_{k=1}^{K} \gamma^p_k \cdot \tilde{X}_k(t) + \sum_{q=1}^{P} \sum_{j=1}^{M} \beta^{pq}_{ij} \cdot \tilde{\lambda}^q(t|j). \quad (21)$$

**Accommodating multiple exogenous signals $\{S_i(t)\}$**

In our VEVO case study (main text Section 6), we consider a different exogenous signal per artist $i$, given by the Google Trends time series $\{S_i(t)\}$. This leads to changes in main text Eq. (5) and main text Eq. (10), since these equations are formulated with a single artist-independent $S(t)$.

Adapting main text Eq. (5) to the case of multiple exogenous signals $\{S_i(t)\}$, we have

$$\lambda^p(t) = \sum_i \mu^p_i \cdot S_i(t) + \sum_{q=1}^{P} \sum_{s<t} \alpha^{pq} \cdot f(t-s) \cdot N^q(s).$$

Adapting main text Eq. (10), we have

$$\lambda^p(t|j) = \mu^p_j \cdot S_j(t) + \sum_{q=1}^{P} \sum_{s<t} \alpha^{pq} \cdot f(t-s) \cdot N^q_j(s).$$

These modifications lead to changes in the structure of the two-tier optimization developed in main text Section 3, since the first-tier parameter set $\Theta_1$ (originally $\{\mu^p, \alpha^{pq}, \theta\}$) now has to include artist-specific parameters $\{\mu^p_i\}$ due to the new form of $\lambda^p(t)$ above. Our new first-tier parameter set $\Theta_1$ becomes $\{\mu^p_i, \alpha^{pq}, \theta\}$, while the new second-tier parameter set $\Theta_2$ becomes $\{\gamma^p_{ik}, \beta^{pq}_{ij}\}$. We add an L2 regularizer on the first-tier optimization and an L1 regularizer on the second-tier optimization to prevent overfitting; we set the regularization parameters to 100.

Furthermore, our gradients also change. For the first-tier likelihood gradients, Eqs. (5) and (7) to (9) are still valid, and we replace Eq. (6) with

$$\partial_{\mu^q_i} \lambda^p(t) = \delta_{pq} \cdot S_i(t),$$

since we now estimate $\{\mu^p_i\}$ instead of $\{\mu^p\}$ in the first-tier optimization.

For the second-tier likelihood gradients, Eqs. (14) and (16) to (20) are still valid, but we do not anymore use Eq. (15) since we do not estimate $\{\mu^p_i\}$ in the second tier.

## E    OMM Fits and Predictions on VEVO 2017 Top 10

Fig. 3 shows the fit and prediction of OMM on the VEVO 2017 Top 10 dataset for the first tier (attention volumes) and second tier (opinion market shares).

## F    Model Elasticities

**Intervention elasticities $e(s_i^p(t), \bar{X}_k(t))$**

Applying main text Eq. (4) on main text Eq. (7) and Eq. (21), we obtain the intervention elasticities $e(s_i^p(t), \bar{X}_k(t))$ as fol-

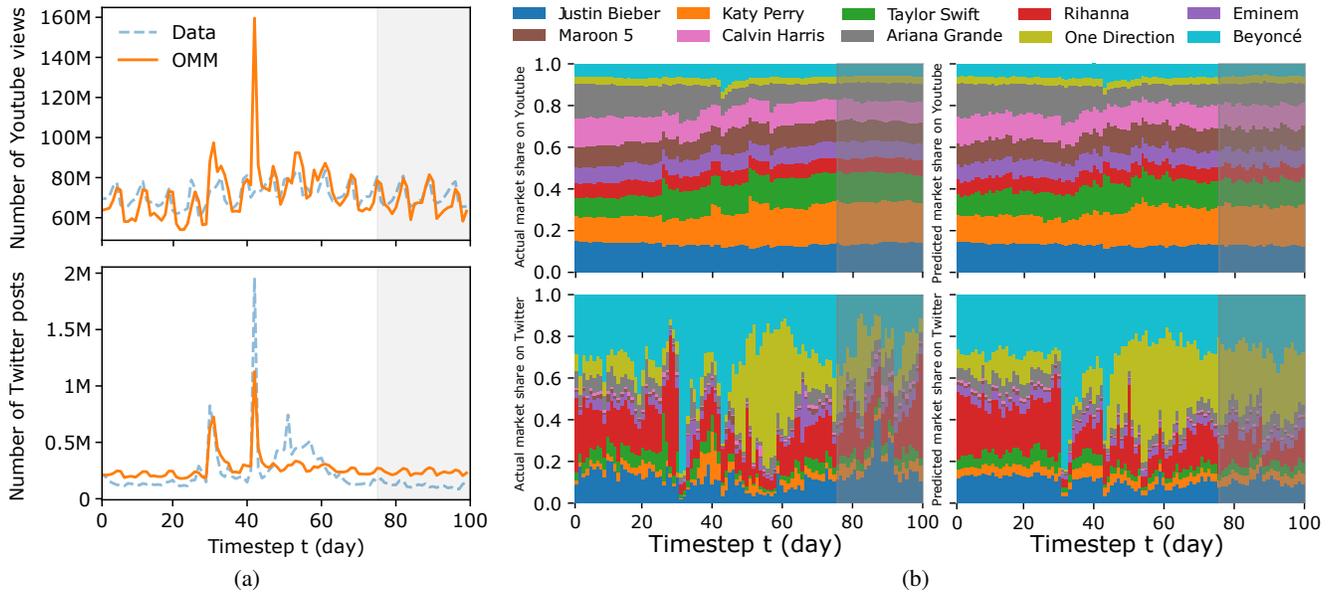

Figure 3: Fitting and predicting with OMM on the VEVO 2017 Top 10 dataset. We train OMM on the first 75 days and predict on days 76 to 100 (shaded area). We show results for Youtube and Twitter, respectively. (a) Actual (dashed blue lines) vs. fitted/predicted (orange lines) volumes; (b) Actual (left panels) and fitted/predicted (right panels) opinion market shares on Youtube (top panels) and Twitter (bottom panels)

lows.

$$\begin{aligned}
&e(s_i^p(t), \bar{X}_k(t)) \\
&= \partial_{\bar{X}_k(t)} s_i^p(t) \cdot \frac{\bar{X}_k(t)}{s_i^p(t)} \\
&= \left\{ -\frac{\mathcal{A}_i^p(t)}{\left[\sum_j \mathcal{A}_j^p(t)\right]^2} \partial_{\bar{X}_k(t)} \sum_j \mathcal{A}_j^p(t) + \frac{\partial_{\bar{X}_k(t)} \mathcal{A}_i^p(t)}{\sum_j \mathcal{A}_j^p(t)} \right\} \cdot \frac{\bar{X}_k(t)}{s_i^p(t)} \\
&= \left\{ -\frac{s_i^p(t)}{\sum_j \mathcal{A}_j^p(t)} \sum_j \mathcal{A}_j^p(t) \frac{\gamma_{jk}^p}{\sigma_{X,k}} + \frac{\mathcal{A}_i^p(t)}{\sum_j \mathcal{A}_j^p(t)} \cdot \frac{\gamma_{ik}^p}{\sigma_{X,k}} \right\} \cdot \frac{\bar{X}_k(t)}{s_i^p(t)} \\
&= \left\{ -\frac{s_i^p(t)}{\sum_j \mathcal{A}_j^p(t)} \sum_j \mathcal{A}_j^p(t) \frac{\gamma_{jk}^p}{\sigma_{X,k}} + s_i^p(t) \cdot \frac{\gamma_{ik}^p}{\sigma_{X,k}} \right\} \cdot \frac{\bar{X}_k(t)}{s_i^p(t)} \\
&= \left\{ -s_i^p(t) \sum_j \left[\frac{\mathcal{A}_j^p(t)}{\sum_l \mathcal{A}_l^p(t)}\right] \frac{\gamma_{jk}^p}{\sigma_{X,k}} + s_i^p(t) \cdot \frac{\gamma_{ik}^p}{\sigma_{X,k}} \right\} \cdot \frac{\bar{X}_k(t)}{s_i^p(t)} \\
&= \left\{ -s_i^p(t) \sum_j s_j^p(t) \frac{\gamma_{jk}^p}{\sigma_{X,k}} + s_i^p(t) \cdot \frac{\gamma_{ik}^p}{\sigma_{X,k}} \right\} \cdot \frac{\bar{X}_k(t)}{s_i^p(t)} \\
&= \left\{ -\sum_j s_j^p(t) \frac{\gamma_{jk}^p}{\sigma_{X,k}} + \frac{\gamma_{ik}^p}{\sigma_{X,k}} \right\} \cdot \bar{X}_k(t) \\
&= \frac{\bar{X}_k(t)}{\sigma_{X,k}} \cdot \sum_j \left[\delta_{ij} - s_j^p(t)\right] \gamma_{jk}^p
\end{aligned}$$

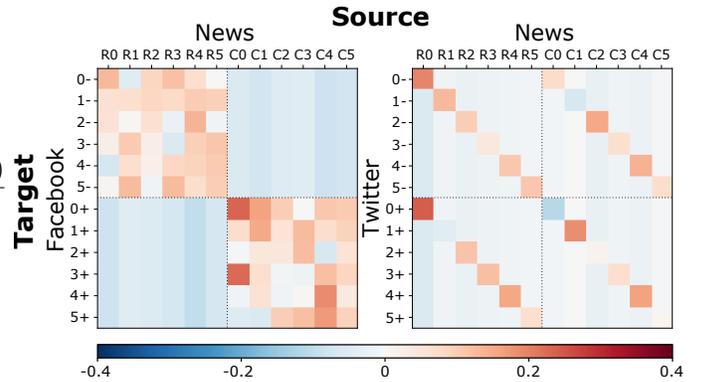

Figure 4: Time-averaged intervention elasticities $e(s_i^p(t), \bar{X}_k(t))$ for the bushfire case study. Elasticities have direction and should be read from column (source) to row (target). The matrix on the left (right) corresponds to influences from reputable (R) and controversial (C) news for each opinion (in $\{0, 1, 2, 3, 4, 5\}$) on the different stanced opinions on Facebook (Twitter).

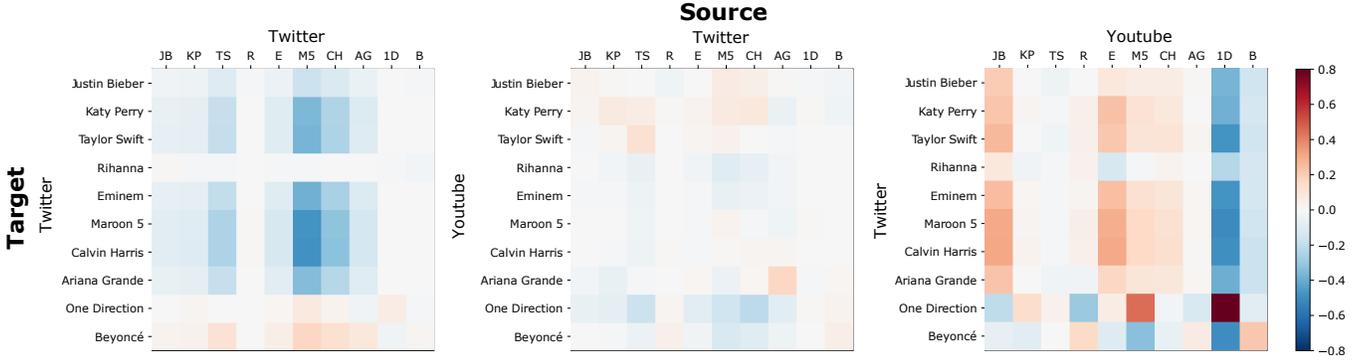

Figure 5: Time-averaged endogenous elasticities $e(s_i^p(t), \lambda^q(t|j))$ of OMM in the VEVO case study. (Left) Twitter-to-Twitter elasticities. (Middle) Twitter-to-Youtube elasticities. (Right) Youtube-to-Twitter elasticities. Elasticities have direction and should be read from column (source) to row (target), both for the platform and within each color matrix.

Time-averaged intervention elasticities for the bushfire case study are shown in Fig. 4.

**Endogenous elasticities** $e(s_i^p(t), \lambda^q(t|j))$

Applying main text Eq. (4) on main text Eq. (7) and Eq. (21), we obtain the endogenous elasticities $e(s_i^p(t), \lambda^q(t|j))$ as follows. Let

$$\phi_j^q(t) = \frac{1}{\operatorname{std}_s \log\left[N_j^q(s)+1\right]} \cdot \frac{1}{\lambda^q(t|j)+1}.$$

We have:

$$
\begin{aligned}
& e(s_i^p(t), \lambda^q(t|j)) \\
&= \partial_{\lambda^q(t|j)} s_i^p(t) \cdot \frac{\lambda^q(t|j)}{s_i^p(t)} \\
&= \left\{ -\frac{\mathcal{A}_i^p(t)}{\left[\sum_k \mathcal{A}_k^p(t)\right]^2} \partial_{\lambda^q(t|j)} \sum_k \mathcal{A}_k^p(t) + \frac{\partial_{\lambda^q(t|j)} \mathcal{A}_i^p(t)}{\sum_k \mathcal{A}_k^p(t)} \right\} \cdot \frac{\lambda^q(t|j)}{s_i^p(t)} \\
&= \left\{ -\frac{s_i^p(t)}{\sum_k \mathcal{A}_k^p(t)} \sum_k \mathcal{A}_k^p(t) \frac{\beta_{kj}^{pq}}{\phi_j^q(t)} + \frac{\mathcal{A}_i^p(t)}{\sum_k \mathcal{A}_k^p(t)} \cdot \frac{\beta_{ij}^{pq}}{\phi_j^q(t)} \right\} \cdot \frac{\lambda^q(t|j)}{s_i^p(t)} \\
&= \left\{ -\frac{s_i^p(t)}{\sum_k \mathcal{A}_k^p(t)} \sum_k \mathcal{A}_k^p(t) \frac{\beta_{kj}^{pq}}{\phi_j^q(t)} + s_i^p(t) \cdot \frac{\beta_{ij}^{pq}}{\phi_j^q(t)} \right\} \cdot \frac{\lambda^q(t|j)}{s_i^p(t)} \\
&= \left\{ -s_i^p(t) \sum_k \left[\frac{\mathcal{A}_k^p(t)}{\sum_l \mathcal{A}_l^p(t)}\right] \frac{\beta_{kj}^{pq}}{\phi_j^q(t)} + s_i^p(t) \cdot \frac{\beta_{ij}^{pq}}{\phi_j^q(t)} \right\} \cdot \frac{\lambda^q(t|j)}{s_i^p(t)} \\
&= \left\{ -s_i^p(t) \sum_k s_k^p(t) \frac{\beta_{kj}^{pq}}{\phi_j^q(t)} + s_i^p(t) \cdot \frac{\beta_{ij}^{pq}}{\phi_j^q(t)} \right\} \cdot \frac{\lambda^q(t|j)}{s_i^p(t)} \\
&= \left\{ -\sum_k s_k^p(t) \frac{\beta_{kj}^{pq}}{\phi_j^q(t)} + \frac{\beta_{ij}^{pq}}{\phi_j^q(t)} \right\} \cdot \lambda^q(t|j) \\
&= \frac{\lambda^q(t|j)}{\phi_j^q(t)} \cdot \sum_k \left[\delta_{ki} - s_k^p(t)\right] \beta_{kj}^{pq}
\end{aligned}
$$

Time-averaged endogenous Twitter-to-Twitter and cross-platform elasticities for the VEVO case study are shown in Fig. 5.

## G  Additional Details on the Bushfire Opinions Dataset

**Dataset construction**

The *Bushfire Opinions dataset* consists of Twitter posts and Facebook posts & comments from Australian user accounts and pages expressing problematic opinions on climate change and the 2019-2020 Australian bushfire season during the 90-day period of November 1, 2019 to January 29, 2020.

The Bushfire Opinions dataset derives from the *SocialSense dataset* introduced in (Kong et al. 2022), which consists of user posts and comments from three major online social media platforms: Facebook, Twitter and Youtube. Postings included in the SocialSense were on two general topics – first, the Australian bushfires and climate change, and second, Covid-19 and vaccination – and expressed problematic opinions. In this work, we focus on Facebook/ Twitter and the Australian bushfires/ climate change topic. Postings were collected using Crowdtangle focused on a set of far-right Australian Facebook groups identified with a digital ethnographic study (for Facebook), the Twitter commercial API (for Twitter), and the Youtube API (for Youtube) using the following keywords as input: *bushfire, australian fires, arson, scottyfrommarketing, liarfromtheshiar, australiaburns, australiaburning, itsthegreensfault, backburning, back burning, climate change, climate emergency, climate hoax, climate crisis, climate action now*. It is important to point out that the Facebook sample is sourced predominantly from far-right groups, whereas the Twitter and Youtube are general scrapes. Two sets of augmentations were added to the postings: the *topic* and the *opinion* of the post, obtained using a set of topic and opinion classifiers trained in (Kong et al. 2022). The set of opinions were constructed via a qualitative study.

A limitation of the original SocialSense dataset is that the Twitter dataset for the Australian bushfires/ climate change topic was scraped only from December 2019 to February 2020, which did not capture early opinion during the start of the bushfire crisis. To that end, we decided to rescrape the Twitter dataset from November 1, 2019 to January 29,

2020 using the Twitter Academic v2 API and the same set of keywords. Since the Twitter Academic API does not allow querying based on user account location, we utilized AWS's Amazon Location Service to geocode users based on their free-text location and description fields and filtered only for tweets from Australian users. Finally, we applied the same set of topic and opinion classifiers to augment the Twitter data.

Once we aligned the Facebook dataset from SocialSense and the rescraped Twitter dataset on the target timeframe, we observed that 10 (out of 34) opinions account for most of the Twitter (95%) and Facebook (81%) postings. To limit the set of opinions in our analysis, we focus on six *opinions of interest* constructed by merging subsets of the 10 opinions, after which we then filter the Twitter and Facebook datasets on this set of opinions. We index the six opinions we consider as $\{0, 1, 2, 3, 4, 5\}$ and are shown below:

0. Greens influence and policy are the cause of the 2019-2020 Australian bushfires./ I am opposed to the policies of Greens political parties.
1. Mainstream media cannot be trusted.
2. Climate change crisis isn't real/ Climate change is a UN hoax/ Climate change is a scam to generate profit for the wealthy and powerful.
3. 2019-2020 Australian bushfires and climate change not related.
4. 2019-2020 Australian bushfires were caused by random arsonists.
5. Changes in the earth's climate are a natural, normal phenomenon/Bush fires are a normal summer occurrence for Australia.

Lastly, keeping in mind our goal of uncovering the interactions between sympathisers and opponents of the aforementioned problematic opinions, we furthermore differentiate whether the expressed opinion shows a *far-right* or *moderate* stance, which effectively splits our set of 6 opinions into 12 *stanced* opinions. For instance, the anti-Greens opinion (labeled 0) splits as far-right (labeled 0+) and moderate (labeled 0+). We represent our set of opinions as $\{(i-, i+)|i \in \{0, \ldots, 5\}\}$. We leverage the far-right stance detector introduced by Ram and Rizoiu (2022) and apply it on each post of the aligned Facebook and Twitter dataset.

In summary, the *Bushfire Opinions dataset* consists of posts on $P = 2$ platforms: 474,461 on Twitter and 27,974 on Facebook, exhibiting $M = 12$ stanced opinions. For compatibility with our discrete-time model, we aggregate post volumes on Facebook and Twitter into hourly counts, yielding $T = 2,160$ time points over the 90-day period of November 1, 2019 to January 29, 2020.



\end{document}